\PassOptionsToPackage{square,sort,comma,numbers}{natbib}

\documentclass{article} 

\usepackage[preprint]{nips_2018}

\usepackage{amsmath}
\usepackage{algorithm}
\usepackage[noend]{algpseudocode}
\usepackage{microtype}
\usepackage{balance}
\usepackage{graphicx}
\usepackage{psfrag}
\usepackage{color}
\usepackage{multirow}
\usepackage{pgfplots}
\usepackage{amsthm, amsmath, amssymb, amsfonts, graphics}
\usepackage{epstopdf, psfrag, pstricks}
\usepackage{booktabs}

\title{Audiovisual Speech Synthesis using Tacotron2}

\author{
Ahmed Hussen Abdelaziz*\\
Apple\\
Cupertino, CA\\
\texttt{ahussenabdelaziz@apple.com}
\And
Anushree Prasanna~Kumar*\\
Apple\\
Cupertino, CA\\
\texttt{ak\_26@apple.com}
\And
Chloe Seivwright\\
Apple\\
London\\
\texttt{cseivwright@apple.com}
\And
Gabriele Fanelli\\
Apple\\
Zurich\\
\texttt{gabriele\_fanelli@apple.com}
\And
Justin Binder\\
Apple\\
Cupertino, CA\\
\texttt{jbinder@apple.com}
\And
Yannis Stylianou\\
Apple\\
London\\
\texttt{istylianou@apple.com}
\And
Sachin Kajarekar\\
Apple\\
Cupertino, CA\\
\texttt{skajarekar@apple.com}
\thanks{Ahmed~H.~Abdelaziz and Anushree~P.~Kumar have contributed equally}
}

\begin{document}

\maketitle

\begin{abstract}
Audiovisual speech synthesis involves synthesizing a talking face while maximizing the coherency of the acoustic and visual speech. To solve this problem, we propose using AVTacotron2, which is an end-to-end text-to-audiovisual speech synthesizer based on the Tacotron2 architecture. AVTacotron2  converts a sequence of pho\-nemes  into a sequence of acoustic features and the corresponding controllers of a face model. The output acoustic features are passed through a WaveRNN model to reconstruct the speech waveform. The speech waveform and the output facial controllers are used to generate the corresponding video of the talking face. As a baseline, we use a modular system, where acoustic speech is synthesized from text using the traditional Tacotron2. The reconstructed acoustic speech is then used to drive the controls of the face model using an independently trained audio-to-facial-animation neural network. We further condition both the end-to-end and modular approaches on emotion embeddings that encode the required prosody to generate emotional audiovisual speech. A comprehensive analysis shows that the end-to-end system is able to synthesize close to human-like audiovisual speech with mean opinion scores (MOS) of 4.1, which is the same MOS obtained on the ground truth generated from professionally recorded videos. 
\end{abstract} 

\noindent{\textbf{Keywords:} Audiovisual speech, speech synthesis, Tacotron2, emotional speech synthesis, blendshape coefficients}

\newpage

\begin{figure*}[t!]
\centering
\def\svgwidth{\linewidth}
\begingroup%
  \makeatletter%
  \providecommand\color[2][]{%
    \errmessage{(Inkscape) Color is used for the text in Inkscape, but the package 'color.sty' is not loaded}%
    \renewcommand\color[2][]{}%
  }%
  \providecommand\transparent[1]{%
    \errmessage{(Inkscape) Transparency is used (non-zero) for the text in Inkscape, but the package 'transparent.sty' is not loaded}%
    \renewcommand\transparent[1]{}%
  }%
  \providecommand\rotatebox[2]{#2}%
  \newcommand*\fsize{\dimexpr\f@size pt\relax}%
  \newcommand*\lineheight[1]{\fontsize{\fsize}{#1\fsize}\selectfont}%
  \ifx\svgwidth\undefined%
    \setlength{\unitlength}{901.47530248bp}%
    \ifx\svgscale\undefined%
      \relax%
    \else%
      \setlength{\unitlength}{\unitlength * \real{\svgscale}}%
    \fi%
  \else%
    \setlength{\unitlength}{\svgwidth}%
  \fi%
  \global\let\svgwidth\undefined%
  \global\let\svgscale\undefined%
  \makeatother%
  \begin{picture}(1,0.4)%
    \lineheight{1}%
    \setlength\tabcolsep{0pt}%
    \put(0,0){\includegraphics[width=\unitlength]{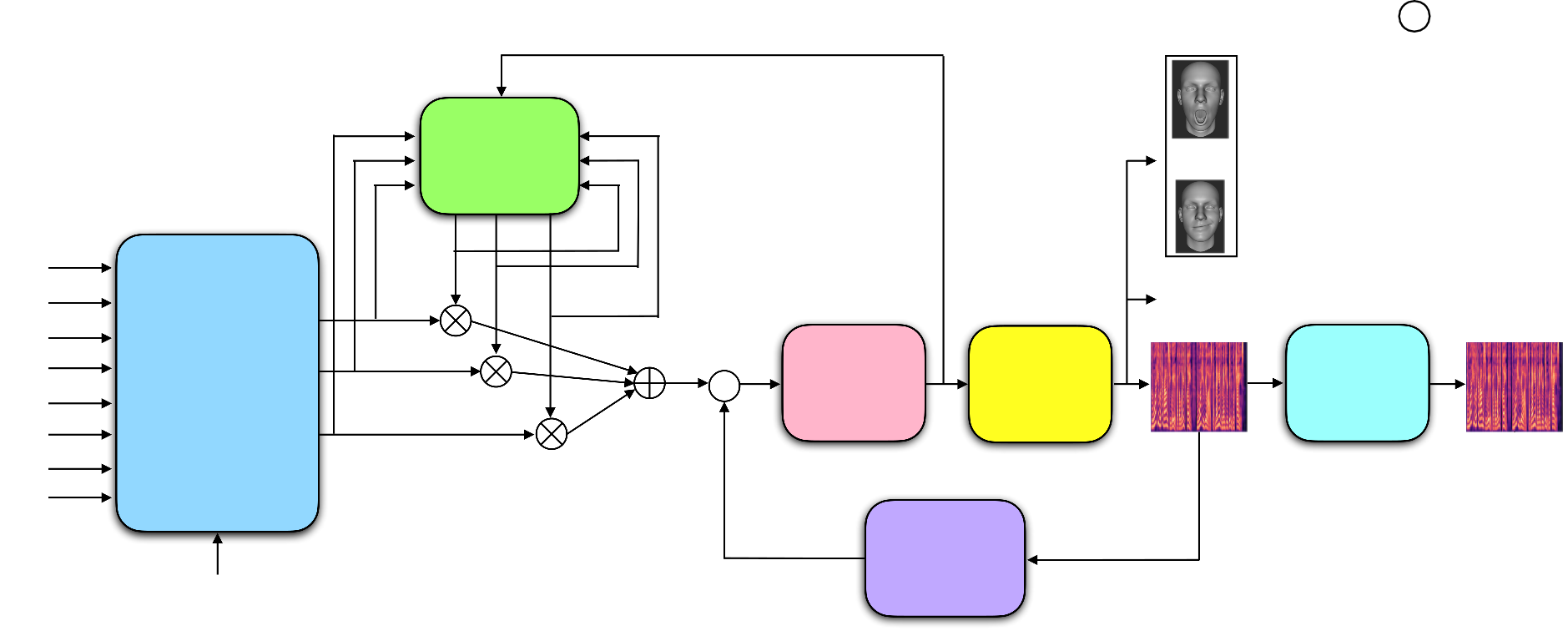}}%
    \put(-0.01,0.22477655){\color[rgb]{0,0,0}\makebox(0,0)[lt]{\lineheight{1.25}\smash{\begin{tabular}[t]{l}\small HH \end{tabular}}}}%
    \put(-0.01,0.20369999){\color[rgb]{0,0,0}\makebox(0,0)[lt]{\lineheight{1.25}\smash{\begin{tabular}[t]{l}\small AH \end{tabular}}}}%
    \put(-0.01,0.18262342){\color[rgb]{0,0,0}\makebox(0,0)[lt]{\lineheight{1.25}\smash{\begin{tabular}[t]{l}\small L \end{tabular}}}}%
    \put(-0.01,0.16154686){\color[rgb]{0,0,0}\makebox(0,0)[lt]{\lineheight{1.25}\smash{\begin{tabular}[t]{l}\small OW\end{tabular}}}}%
    \put(-0.01,0.1404703){\color[rgb]{0,0,0}\makebox(0,0)[lt]{\lineheight{1.25}\smash{\begin{tabular}[t]{l}\small W\end{tabular}}}}%
    \put(-0.01,0.11939374){\color[rgb]{0,0,0}\makebox(0,0)[lt]{\lineheight{1.25}\smash{\begin{tabular}[t]{l}\small ER \end{tabular}}}}%
    \put(-0.01,0.09831718){\color[rgb]{0,0,0}\makebox(0,0)[lt]{\lineheight{1.25}\smash{\begin{tabular}[t]{l}\small L \end{tabular}}}}%
    \put(-0.01,0.07724062){\color[rgb]{0,0,0}\makebox(0,0)[lt]{\lineheight{1.25}\smash{\begin{tabular}[t]{l}\small D\end{tabular}}}}%
    \put(0.10000806,0.15285099){\color[rgb]{0,0,0}\makebox(0,0)[lt]{\lineheight{1.25}\smash{\begin{tabular}[t]{l}\small Encoder\end{tabular}}}}%
    \put(0.22309067,0.13182125){\color[rgb]{0,0,0}\rotatebox{90}{\makebox(0,0)[lt]{\lineheight{1.25}\smash{\begin{tabular}[t]{l}\textbf{\small …}\end{tabular}}}}}%
    \put(0.32154815,0.22110295){\color[rgb]{0,0,0}\makebox(0,0)[lt]{\lineheight{1.25}\smash{\begin{tabular}[t]{l}\textbf{\small …}\end{tabular}}}}%
    \put(0.36702106,0.20288487){\color[rgb]{0,0,0}\rotatebox{90}{\makebox(0,0)[lt]{\lineheight{1.25}\smash{\begin{tabular}[t]{l}\textbf{\small …}\end{tabular}}}}}%
    \put(0.275,0.29764341){\color[rgb]{0,0,0}\makebox(0,0)[lt]{\lineheight{1.25}\smash{\begin{tabular}[t]{l}\small Attention\end{tabular}}}}%
    \put(0.505,0.15210887){\color[rgb]{0,0,0}\makebox(0,0)[lt]{\lineheight{1.25}\smash{\begin{tabular}[t]{l}\small Decoder\end{tabular}}}}%
    \put(0.62,0.15210887){\color[rgb]{0,0,0}\makebox(0,0)[lt]{\lineheight{1.25}\smash{\begin{tabular}[t]{l}\small Regressor\end{tabular}}}}%
    \put(0.835,0.15285099){\color[rgb]{0,0,0}\makebox(0,0)[lt]{\lineheight{1.25}\smash{\begin{tabular}[t]{l}\small Postnet\end{tabular}}}}%
    \put(0.575,0.0410719){\color[rgb]{0,0,0}\makebox(0,0)[lt]{\lineheight{1.25}\smash{\begin{tabular}[t]{l}\small Prenet\end{tabular}}}}%
    \put(0.45653077,0.15245045){\color[rgb]{0,0,0}\makebox(0,0)[lt]{\lineheight{1.25}\smash{\begin{tabular}[t]{l}\small =\end{tabular}}}}%
    \put(0.45653077,0.14864077){\color[rgb]{0,0,0}\makebox(0,0)[lt]{\lineheight{1.25}\smash{\begin{tabular}[t]{l}\small =\end{tabular}}}}%
    \put(0.76751015,0.29){\color[rgb]{0,0,0}\rotatebox{90}{\makebox(0,0)[lt]{\lineheight{1.25}\smash{\begin{tabular}[t]{l}\small …\end{tabular}}}}}%
    \put(0.89684444,0.38827664){\color[rgb]{0,0,0}\makebox(0,0)[lt]{\lineheight{1.25}\smash{\begin{tabular}[t]{l}\small =\end{tabular}}}}%
    \put(0.89684444,0.38446696){\color[rgb]{0,0,0}\makebox(0,0)[lt]{\lineheight{1.25}\smash{\begin{tabular}[t]{l}\small =\end{tabular}}}}%
    \put(0.92330653,0.3864586){\color[rgb]{0,0,0}\makebox(0,0)[lt]{\lineheight{1.25}\smash{\begin{tabular}[t]{l}\small Concatenation\end{tabular}}}}%
    \put(0.03,0.01950293){\color[rgb]{0,0,0}\makebox(0,0)[lt]{\lineheight{1.25}\smash{\begin{tabular}[t]{l}\small Emotion embeddings\end{tabular}}}}%
    \put(0.91954314,0.10590951){\color[rgb]{0,0,0}\makebox(0,0)[lt]{\lineheight{1.25}\smash{\begin{tabular}[t]{l}\small Mel-scaled\end{tabular}}}}%
    \put(0.92101202,0.088){\color[rgb]{0,0,0}\makebox(0,0)[lt]{\lineheight{1.25}\smash{\begin{tabular}[t]{l}\small filterbanks\end{tabular}}}}%
    \put(0.72408702,0.39251011){\color[rgb]{0,0,0}\makebox(0,0)[lt]{\lineheight{1.25}\smash{\begin{tabular}[t]{l}\small Blendshape\end{tabular}}}}%
    \put(0.72509069,0.373){\color[rgb]{0,0,0}\makebox(0,0)[lt]{\lineheight{1.25}\smash{\begin{tabular}[t]{l}\small coeffcients\end{tabular}}}}%
    \put(0.74359841,0.20728464){\color[rgb]{0,0,0}\makebox(0,0)[lt]{\lineheight{1.25}\smash{\begin{tabular}[t]{l}\small Endpointer\end{tabular}}}}%
  \end{picture}%
\endgroup%
\caption{AVTacotron2 architecture for emotional audiovisual speech synthesis}
\label{fig:framework}
\end{figure*}
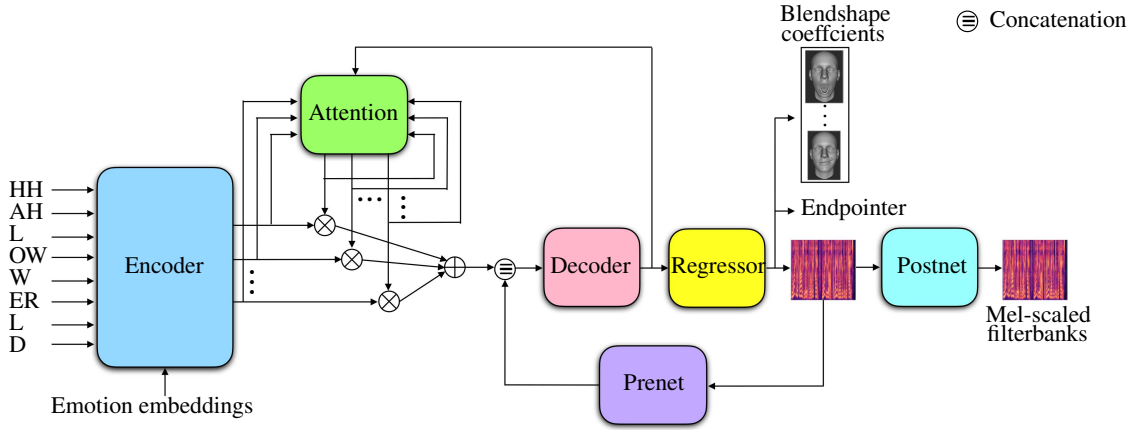

\section{Introduction} \label{sec:intro}

Human perception of speech is inherently bimodal. In face-to-face conversations, vision improves the intelligibility of speech by reading the lips of the talker. Such forms of visual hearing are exploited not only by hearing impaired people, but also by all individuals \cite{cotton1935normal}. The presumed biological link between perception and production makes bimodal speech production essential in generating talking characters. Thus, audiovisual speech synthesis involves synthesizing visual controllers for a face model that is coherent with the synthesized acoustic speech.  This is extremely challenging, as humans are highly sensitive to subtle discrepancies between lip movements and accompanying sound. Poor visual speech synthesis can negatively influence the intelligibility of what is being spoken \cite{mcgurk1976hearing} even if the synthesized acoustic speech is natural.  Additionally, since speech can be spoken in different ways depending on the emotional state, it is essential to synthesize plausible facial expressions along with acoustic speech and lip movements to increase the overall naturalness.

Recent advances in neural text-to-speech (TTS) systems have led to synthesizing close-to-human-like speech. One of the widely used neural TTS architectures is Tacotron 2 \cite{shen2018natural}. In this paper, we extend Tacotron 2 for the task of audiovisual speech synthesis. The audiovisual Tactoron2 (AVTacotron2), is an end-to-end approach, where facial controls along with the acoustic spectral features are synthesized directly from text. The resulting spectral features are used to reconstruct the speech signal using WaveRNN \cite{kalchbrenner2018efficient}. The facial controllers generated by the AVTacotron2 are used to animate a 3D face model. Finally, the video of the talking face is generated by combining the synthesized acoustic and visual information. We also use Tacotron2 in our baseline to synthesize acoustic speech from text. An independent speech-to-animation module is then used to generate the corresponding facial controlsh. Speech signal reconstruction and video generation are done in the same way as in the end-to-end approach. 

In this paper we show that the proposed single neural network captures the correlation between the audio and visual cues whilst preserving the synergy between the synthesized acoustic and visual speech. AVTacotron2 synthesizes talking faces that can be directly deployed in computer graphics applications, such as computer-generated movies, video games, video conferencing, and virtual agents, without any manual post--processing. Our results show that AVTacotron2 gives better overall synthesis quality compared to the modular baseline system.


We represent the space of facial motion using generic blendshapes. The benefits of using such a facial representation are: 1) The controls have  semantic meaning which makes analyzing and post-\\processing them for further improvements easy. 2) The blendshape coefficients are independent of the face model and can be applied to a variety of fictional and human-like characters.

Modulating synthesized speech with the corresponding prosody \cite{skerry2018towards} and facial expressions is a crucial for audiovisual speech synthesis. Recently, significant breakthroughs in style modeling have been proposed in speech synthesis \cite{wang2018style}. In this study, we address the synthesis of easy-to-define styles of speech, which are emotions. We present a supervised approach for emotion-aware audiovisual speech synthesis. In particular, we condition both the end-to-end and modular approaches on emotion embeddings that encode the required prosody of audiovisual emotional speech. The emotion embeddings are extracted from a stand alone speech emotion classifier.  

Evaluating audio-only and audiovisual speech synthesis systems need to be assessed using human perception. In this work, we evaluate the two proposed systems using subjective tests. The tests focus on different aspects of the synthesis, such as the quality of acoustic speech, lip movement, facial expression, and emotion in speech. We use mean opinion scores (MOS) tests and AB tests to obtain absolute and relative scores for the two approaches. We also evaluate the holistic perception of the talking face.

To summarize, our main contributions are:
\begin{itemize}
\item{We propose and evaluate AVTacotron2, an end-to-end system for emotion-aware audiovisual speech synthesis. We show that AVTacotron2 generates close to human-like emotional audiovisual speech without post-processing.}
\item{We apply AVTacotron2 to a generic blendshape model that allows for synthesizing talking faces for a variety of face models.}
\item{We perform comprehensive empirical evaluation to assess the quality of different aspects of the synthesized speech as well as the overall quality. We compare the synthesis quality of the proposed systems to the original recordings of a professional actor.}
\end{itemize}

The rest of the paper is organized as follows:  In Section~\ref{sec:rw}, we discuss related work. Section~\ref{sec:bsc} describes offline extraction of facial controls from video sequences. In Section~\ref{sec:emotion_recog}, we introduce the speech emotion recognition system used for extracting emotion embeddings. Section~\ref{tta} outlines the model architectures of AVTacotron2 for synthesizing audiovisual speech from text.  The baseline modular approach is then described in Section~\ref{tta_mod}. The experiments used to evaluate the model performance and the results are described in Section~\ref{sec:results}. Finally, we conclude the paper and give an outline of future work in Section~\ref{sec:conclsion}. 

\section{Related work} \label{sec:rw}

\textbf{Manually and Video-driven Avatars:} Despite improvements in sophisticated face and motion tracking systems, the cost of production remains high. In many production houses, high fidelity speech animation is either created manually by an animator or by using facial motion capture of human actors \cite{beeler2011high,cao2015real,fyffe2013driving,weng2014real,zhang2008spacetime}. 

\textbf{Face Representations:} Face synthesis algorithms can be categorized into two major categories: photorealistic and parametric. In photorealistic synthesis, videos of the speakers are generated either from text or audio inputs \cite{linsen2020ebt,kumar2017obamanet,thies2019nvp}. These techniques involve combining both image and speech synthesis tasks. 

In parametric face synthesis, a sequence of facial controllers are used to deform a neutral face model. The parametric approaches can either operate on 3D face models \cite{anderson2013expressive,cao2016real} or on 2D images. For the 2D face synthesis, active appearance models (AAMs) \cite{cootes2001active} are widely used.

\textbf{Text-driven Avatars:} Many approaches have been proposed for visual and audiovisual speech synthesis from text. For example, hidden Markov models (HMM) have been used \cite{sako2000hmm,xie2007realistic,fu2005audio,wang2008real}. Decision trees were used to create expressive talking heads in \cite{anderson2013expressive}. In \cite{goyal2000text}, the Festival System was first used for speech synthesis followed by a text-to-visual-speech synthesis system and an audiovisual synchronizer. In \cite{taylor2012dynamic}, a learned phoneme-to-viseme mapping was used for visual speech synthesis. Recently, many deep-learning-\\based approaches have been proposed. In \cite{parker2017expressive}, the authors used a unit-selection-based audiovisual speech synthesis system where the output of a deep neural network (DNN) was used as a likelihood of each audiovisual unit in an inventory. Dynamic programming provided the most likely trajectory of consecutive audiovisual units, which were later used for synthesis. A more neural based approach was proposed in \cite{taylor2017deep}, where a fully connected feed-forward neural network was used to synthesize natural speech animation that is coherent with the input text and speech. In \cite{kumar2017obamanet}, an LSTM architecture was used to convert input text into speech and corresponding coherent photo-realistic images of talking faces.

\textbf{Speech-driven Avatars:} Voice Puppetry \cite{brand1999voice} is an early HMM-based approach for audio-driven facial animation. This was followed by techniques based on decision trees \cite{kim2015decision}, or deep bidirectional LSTM \cite{fan2015photo} that outperformed \\ HMM-based approaches. In \cite{shimba2015talking}, the authors applied an LSTM-based neural network directly to audio features. In \\ \cite{taylor2016audio}, a deep neural network was proposed to regress to a window of visual features from a sliding window of audio features. In that work, a time-shift recurrent network trained on unlabeled visual speech dataset produced convincing results without dependencies on a speech recognition system. More recently, convolutional neural networks (CNNs) were used in  \cite{karras2017audio} to learn a mapping from speech to the vertex coordinates of a 3D face model. Facial expressions and emotion were generated by a latent code extracted from the network. In \cite{zhou2018visemenet}, the authors propose VisemeNet which consisted of a three-stage LSTM network for lip sync. Most of these systems synthesize visual controllers that deform a speaker dependent 3D face model. 

\textbf{Discussion:} Both manual and video-driven approaches require skilled animators to edit complex animation parameters. This is extremely time consuming and expensive, especially in applications such as computer generated movies and digital game production, where talking faces for tens of hours of dialog are required. As a solution for that, we propose using a novel text-driven approach for synthesizing production-quality speech and facial animation with different styles of speech without the need for post processing.

Our approach belongs to the parametric category of face representations, where a sequence of low-dimensional blendshape coefficients (BSC) are synthesized and applied to a wide range of avatars. We extract ground truth BSCs similarly to \cite{weise2011realtime}, which are used as facial controllers of face models.

Compared to all text-driven facial animation approaches, we propose the AVTacotron2 to synthesize audiovisual speech in an end-to-end neural fashion to generate a talking face.

Our modular baseline has a speech-driven animation module, which uses a CNN-based model similar to \cite{karras2017audio}, but in contrast, we predict generic BSCs that can be used across various fictional and human-like characters.

\section{Audiovisual Tacotron2}

\subsection{Estimation of the Ground Truth Blendshape Coefficients (BSC)}\label{sec:bsc}

A challenging problem in facial animation is to find the ground-truth controllers of a face model. Manual labeling of the facial controls for each frame in a dataset that is large enough to train a neural network is impractical. Annotation is both subjective, time consuming and expensive. In this study, we follow the method described in~\cite{abdelaziz2020modality} to automatically estimate the facial controls from RGB-D video streams.

\subsection{Speech Emotion Features} \label{sec:emotion_recog}

Emotions in visual speech synthesis systems are increasingly required since they enhance the user experience and make the interaction more natural. Adding emotions to the synthesized acoustic and visual speech requires conditioning a model on a representation of the required emotion. In contrast to the style transfer approaches in \cite{wang2018style} and \cite{skerry2018towards} that have been recently successful in learning the speech style/emotion in an unsupervised manner, we extract the emotion representations from an emotion classifier that was trained using available emotion labels. Our supervised approach gives empirically comparative results to the unsupervised approach and allows finer control on the level of the synthesized emotion.  We use the penultimate layer of a speech emotion classifier as a fine-grained vector representation of the emotion. Three emotion categories are investigated: neutral, positive (happy) and negative (sad).

\begin{figure}[t!]
\centering
\def\svgwidth{0.2\linewidth}
\begingroup%
  \makeatletter%
  \providecommand\color[2][]{%
    \errmessage{(Inkscape) Color is used for the text in Inkscape, but the package 'color.sty' is not loaded}%
    \renewcommand\color[2][]{}%
  }%
  \providecommand\transparent[1]{%
    \errmessage{(Inkscape) Transparency is used (non-zero) for the text in Inkscape, but the package 'transparent.sty' is not loaded}%
    \renewcommand\transparent[1]{}%
  }%
  \providecommand\rotatebox[2]{#2}%
  \newcommand*\fsize{\dimexpr\f@size pt\relax}%
  \newcommand*\lineheight[1]{\fontsize{\fsize}{#1\fsize}\selectfont}%
  \ifx\svgwidth\undefined%
    \setlength{\unitlength}{276.30349309bp}%
    \ifx\svgscale\undefined%
      \relax%
    \else%
      \setlength{\unitlength}{\unitlength * \real{\svgscale}}%
    \fi%
  \else%
    \setlength{\unitlength}{\svgwidth}%
  \fi%
  \global\let\svgwidth\undefined%
  \global\let\svgscale\undefined%
  \makeatother%
  \begin{picture}(1,3.0)%
    \lineheight{1}%
    \setlength\tabcolsep{0pt}%
    \put(0,0){\includegraphics[width=\unitlength]{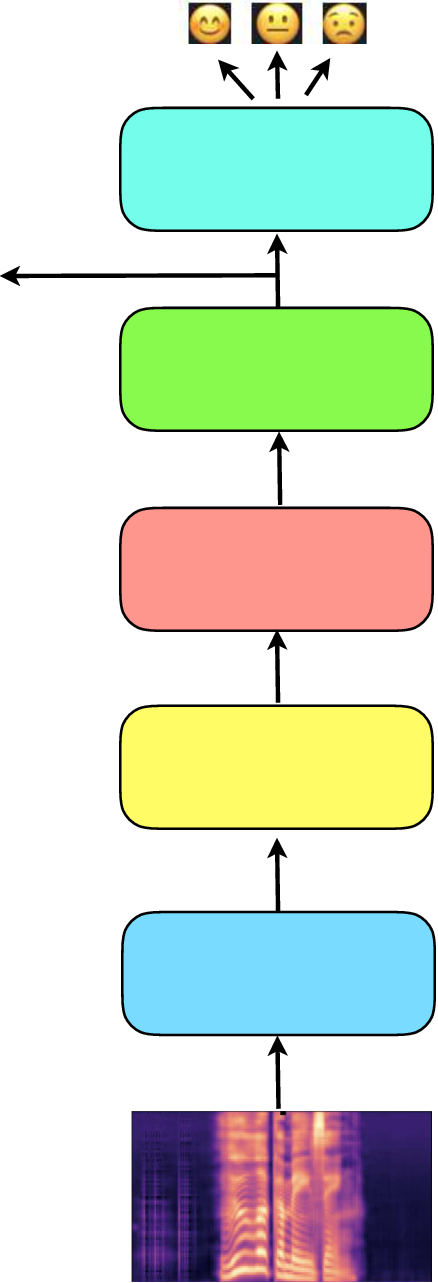}}%
    \put(0.44,0.68){\color[rgb]{0,0,0}\makebox(0,0)[lt]{\lineheight{1.25}\smash{\begin{tabular}[t]{l}{\small Conv1D}\end{tabular}}}}%
    \put(0.47,1.15){\color[rgb]{0,0,0}\makebox(0,0)[lt]{\lineheight{1.25}\smash{\begin{tabular}[t]{l}{\small LSTM}\end{tabular}}}}%
    \put(0.56,1.6){\color[rgb]{0,0,0}\makebox(0,0)[lt]{\lineheight{1.25}\smash{\begin{tabular}[t]{l} {\small FC}\end{tabular}}}}%
    \put(0.33,2.06){\color[rgb]{0,0,0}\makebox(0,0)[lt]{\lineheight{1.25}\smash{\begin{tabular}[t]{l}{\small Average Pool}\end{tabular}}}}%
    \put(0.44,2.52){\color[rgb]{0,0,0}\makebox(0,0)[lt]{\lineheight{1.25}\smash{\begin{tabular}[t]{l}{\small Softmax}\end{tabular}}}}%
    \put(-0.6,2.35){\color[rgb]{0,0,0}\makebox(0,0)[lt]{\lineheight{1.25}\smash{\begin{tabular}[t]{l}{\small Emotion}\end{tabular}}}}%
    \put(-0.68,2.25){\color[rgb]{0,0,0}\makebox(0,0)[lt]{\lineheight{1.25}\smash{\begin{tabular}[t]{l}{\small Embeddings}\end{tabular}}}}%
  \end{picture}%
\endgroup%
\caption{Speech emotion recognition neural network. The network is trained to classify neutral, happy, and sad emotions using input acoustic features. Once the network is trained, the output of the average pooling layer is used as a speech emotion representation.}
\label{fig:speech_emo_net}
\end{figure}
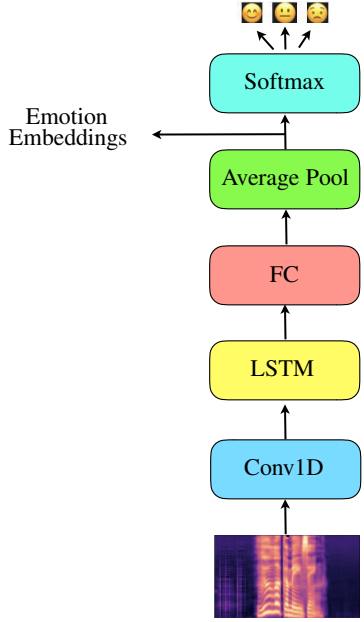

Figure \ref{fig:speech_emo_net} shows the architecture of the speech emotion classifier. In this approach, 40-dimensional mel-scaled filter bank features are first passed through a 1d-convolutional layer, which contains 100 filters with kernel size of $3\times1$. The output of this convolutional layer is then fed into a single long short-term memory (LSTM) layer with 64 cells followed by a linear layer of the same dimension. Mean pooling is then used to summarize the LSTM outputs across time. A softmax layer is applied to the output of the average pooling layer to predict the emotion class.

The output of the penultimate layer of the trained model is used as the speech emotion embedding. These embeddings are used to condition the audiovisual speech synthesis systems during training. During inference, the embedding from a reference utterance is used to get the desired emotion for the synthesized talking face. 

\section{End-to-end Audiovisual Speech Synthesis} \label{tta}

\subsection{End-to-end Audiovisual Speech Synthesis} \label{tta}

Deep neural networks, in particular sequence-to-sequence architectures \cite{wang2017tacotron,shen2018natural}, produce almost human-like synthesized speech. Tacotron2 \cite{shen2018natural} is one of the most successful sequence-to-sequence architectures for speech synthesis. The Tacotron2 architecture converts a phoneme sequence of length $M$ that represents the pronunciation of the sentence to be synthesized to 80-dimensional mel-scaled filter bank (MFB) features. The MFB features are input to another sequence-to-sequence model, e.g., WaveRNN \cite{kalchbrenner2018efficient}, to generate the samples of the synthesized speech signal. 


Figure \ref{fig:framework} shows the AVTacotron2 architecture. Although the network is trained end-to-end, it can be divided into four functional blocks: a phoneme encoder, an attention, a decoder, and a regressor blocks.  In the following sections, these four blocks are described.

\subsubsection{Phoneme Encoder} \label{ttae}

\begin{figure}[t!]
\centering
\def\svgwidth{0.5\linewidth}
\begingroup%
  \makeatletter%
  \providecommand\color[2][]{%
    \errmessage{(Inkscape) Color is used for the text in Inkscape, but the package 'color.sty' is not loaded}%
    \renewcommand\color[2][]{}%
  }%
  \providecommand\transparent[1]{%
    \errmessage{(Inkscape) Transparency is used (non-zero) for the text in Inkscape, but the package 'transparent.sty' is not loaded}%
    \renewcommand\transparent[1]{}%
  }%
  \providecommand\rotatebox[2]{#2}%
  \newcommand*\fsize{\dimexpr\f@size pt\relax}%
  \newcommand*\lineheight[1]{\fontsize{\fsize}{#1\fsize}\selectfont}%
  \ifx\svgwidth\undefined%
    \setlength{\unitlength}{434.26945936bp}%
    \ifx\svgscale\undefined%
      \relax%
    \else%
      \setlength{\unitlength}{\unitlength * \real{\svgscale}}%
    \fi%
  \else%
    \setlength{\unitlength}{\svgwidth}%
  \fi%
  \global\let\svgwidth\undefined%
  \global\let\svgscale\undefined%
  \makeatother%
  \begin{picture}(1.2,1.25)%
    \lineheight{1}%
    \setlength\tabcolsep{0pt}%
    \put(0,0){\includegraphics[width=\unitlength]{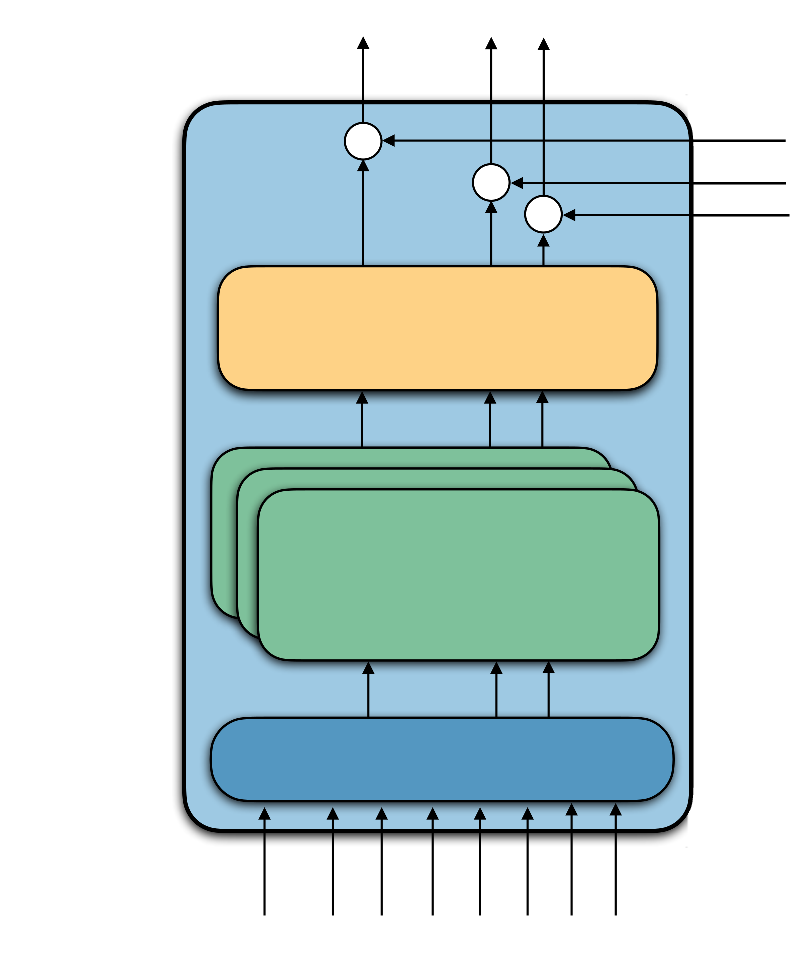}}%
    \put(0.447,1.032){\color[rgb]{0,0,0}\makebox(0,0)[lt]{\lineheight{1.25}\smash{\begin{tabular}[t]{l}=\end{tabular}}}}%
    \put(0.447,1.023){\color[rgb]{0,0,0}\makebox(0,0)[lt]{\lineheight{1.25}\smash{\begin{tabular}[t]{l}=\end{tabular}}}}%
    \put(0.396,0.24){\color[rgb]{0,0,0}\makebox(0,0)[lt]{\lineheight{1.25}\smash{\begin{tabular}[t]{l}Embedding layer\end{tabular}}}}%
    \put(0.476,0.785){\color[rgb]{0,0,0}\makebox(0,0)[lt]{\lineheight{1.25}\smash{\begin{tabular}[t]{l}BLSTM\end{tabular}}}}%
    \put(0.211,1.16){\color[rgb]{0,0,0}\makebox(0,0)[lt]{\lineheight{1.25}\smash{\begin{tabular}[t]{l}Encoder \end{tabular}}}}%
    \put(0.227,1.13){\color[rgb]{0,0,0}\makebox(0,0)[lt]{\lineheight{1.25}\smash{\begin{tabular}[t]{l}output\end{tabular}}}}%
    \put(-0.00046549,0.6061585){\color[rgb]{0,0,0}\makebox(0,0)[lt]{\lineheight{1.25}\smash{\begin{tabular}[t]{l}Encoder\end{tabular}}}}%
    \put(1.04,1.0){\color[rgb]{0,0,0}\makebox(0,0)[lt]{\lineheight{1.25}\smash{\begin{tabular}[t]{l}Emotion \end{tabular}}}}%
    \put(1.02,0.95){\color[rgb]{0,0,0}\makebox(0,0)[lt]{\lineheight{1.25}\smash{\begin{tabular}[t]{l}embedding\end{tabular}}}}%
    \put(0.296,0.00054375){\color[rgb]{0,0,0}\makebox(0,0)[lt]{\lineheight{1.25}\smash{\begin{tabular}[t]{l}{ \footnotesize HH \ AH L OW W ER \ L \ D}\end{tabular}}}}%
    \put(0.516,0.47){\color[rgb]{0,0,0}\makebox(0,0)[lt]{\lineheight{1.25}\smash{\begin{tabular}[t]{l}CNN\end{tabular}}}}%
    \put(0.675,0.941){\color[rgb]{0,0,0}\makebox(0,0)[lt]{\lineheight{1.25}\smash{\begin{tabular}[t]{l}=\end{tabular}}}}%
    \put(0.675,0.932){\color[rgb]{0,0,0}\makebox(0,0)[lt]{\lineheight{1.25}\smash{\begin{tabular}[t]{l}=\end{tabular}}}}%
    \put(0.516,0.9){\color[rgb]{0,0,0}\makebox(0,0)[lt]{\lineheight{1.25}\smash{\begin{tabular}[t]{l}\textbf{…}\end{tabular}}}}%
    \put(0.516,0.34){\color[rgb]{0,0,0}\makebox(0,0)[lt]{\lineheight{1.25}\smash{\begin{tabular}[t]{l}\textbf{…}\end{tabular}}}}%
    \put(0.516,0.68){\color[rgb]{0,0,0}\makebox(0,0)[lt]{\lineheight{1.25}\smash{\begin{tabular}[t]{l}\textbf{…}\end{tabular}}}}%
    \put(0.607,0.98){\color[rgb]{0,0,0}\makebox(0,0)[lt]{\lineheight{1.25}\smash{\begin{tabular}[t]{l}=\end{tabular}}}}%
    \put(0.607,0.971){\color[rgb]{0,0,0}\makebox(0,0)[lt]{\lineheight{1.25}\smash{\begin{tabular}[t]{l}=\end{tabular}}}}%
  \end{picture}%
\endgroup%
\label{fig:encoder}
\caption{Emotion-conditioned phoneme encoder.}
\end{figure}
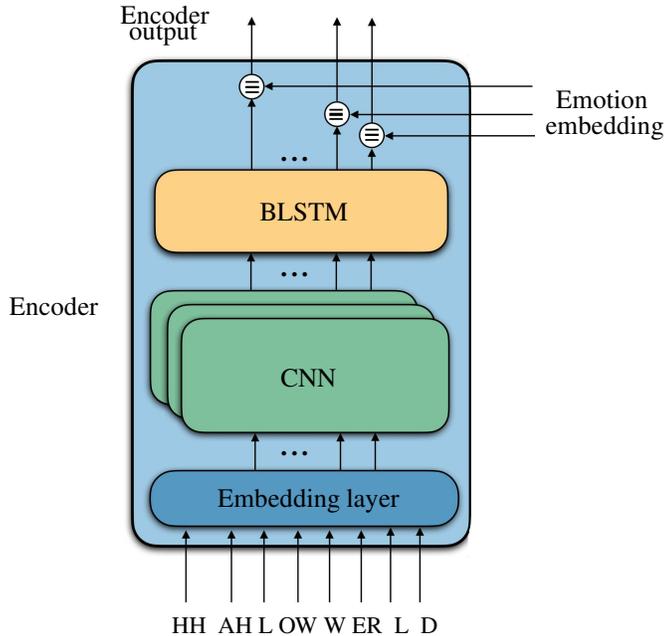

The function of the phoneme encoder in the Tacotron2 architecture is to transform a sequence of phonemes into an intermediate representation, from which acoustic and visual features can be generated. As shown in Figure \ref{fig:encoder}, the input to the encoder is a sequence of $M$ phoneme indices, which are transformed into 512-dimensional feature vectors using an embedding layer as described in \cite{shen2018natural}.  This embedding vector is passed through 3 convolutional layers, each of which has 512 feature maps. The output of the final convolutional layer is processed by a bi-directional long short-term memory (BLSTM) layer with 512 units in each direction. The concatenation of the bi-directional LSTM gives the $M \times1024$ intermediate representation of the $M$ input phonemes. The 64 dimensional emotion embedding is repeated and cascaded with the processed phoneme embeddings to give the encoder outputs. During training the embeddings are extracted from the same output utterance. During inference, the emotion embedding is extracted from a reference utterance that has the required level of emotion.  The emotionally conditioned phoneme representation is used to generate the audio-visual outputs in the later stages of the network, as described in the following sections.


\subsubsection{Attention mechanism} \label{ttaa}

A location-sensitive attention \\ \cite{chorowski2015attention} is applied to the $M \times1088$ intermediate phoneme representation generated by the phoneme encoder. The output of the attention mechanism is called the context vector and is estimated as a weighted sum of the $M$ intermediate representations. The attention and decoding blocks are applied in an iterative way.  In the traditional Tacotron2 model, the attention learns correlations between text (phoneme representation) and audio. In the proposed AVTacotron2, we extend the attention module to learn correlations between text and facial controllers in addition to speech. The attention mechanism receives additional inputs which correspond to the visual features of the previous time step from the decoder. The learned attention weights, i.e. alignments, are processed by a 1D convolutional layer of 32 feature maps, each of 31-dimensions.  The context vector has 1088 dimensions and is used as the input to the decoder network.
%

\subsubsection{Decoder} \label{ttad}

The decoder stage of the Tacotron2 architecture is composed of two 1024-dimensional uni-directional LSTM layers. The decoder iteratively generates an embedding, from which all audio, video, and auxiliary information can be estimated. The input to the decoder at time-frame $t$ is the context vector at time-frame $t$ and a pre-processed version of the regressor output of the previous time-frame $t-1$. Preprocessing of the regressor outputs is done by the so-called Pre-net, which is composed of two 256-dimensional fully-connected layers with ReLU activations.  Since the decoder generates embeddings that are asynchronous with the encoder output, the number of output embeddings $N_\text{decoder}$ is independent of the encoder output length $M$.  The generated embedding for visual features can be split into lower half (lip movements) and upper half (facial expressions). In general, lip movements tend to be better correlated with what was spoken. To maximize the correlation with text, we experimented with two feedback approaches to the attention module --- \textit{full feedback}, where the input from the previous time step consisted of the full decoder output, and \textit{limited feedback}, where the input from the previous time step was restricted to audio and lower half facial features. 

\subsubsection{Regressor} \label{ttar}

The embeddings extracted from the decoder encode all information needed to synthesize one or $n$ acoustic and visual frames. We use $n=2$ in this study, which means that the length of the synthesized acoustic and visual features is $2N_\text{decoder}$. The regressor is simply a linear projection of the embeddings from the decoder output onto the acoustic, the visual, and the auxiliary feature spaces. The acoustic feature space is the 80-dimensional MFBs. The visual feature space is the 51-dimensional BSCs. The auxiliary feature is a one-dimensional end-pointing feature that determines the end of decoding. 

The output MFB features go through a post-processing network (post-net) that is composed of five 1D convolutional layers followed by a linear fully-connected layer. The convolutional layers have 512 filters of kernel size 5. The final fully connected layer has 80 units, which is the same dimension as the regression acoustic feature space. 



The network is trained to minimize a weighted sum of four loss functions. 
\begin{equation}
\mathcal{L}_\text{total} = \omega_\text{mel} \mathcal{L}_\text{mel}  + \omega_\text{post} \mathcal{L}_\text{post} + \omega_\text{end} \mathcal{L}_\text{end} + \omega_\text{bsc} \mathcal{L}_\text{bsc} . 
\label{eq:total}
\end{equation}
$\mathcal{L}_\text{mel}$  is the L1 loss (mean absolute error) applied to the mel-scaled filter-bank features. $\mathcal{L}_\text{post}$ and  $\mathcal{L}_\text{end}$ are the L2 (mean square error) losses that are applied to the output of the post-processing subnetwork and the end-pointing output respectively. For the BSCs, we use the L2 loss function $\mathcal{L}_\text{bsc}$. The weights $ \omega_\text{mel},  \omega_\text{post}, \omega_\text{end}, \text{and}  \omega_\text{bsc}$ are empirically tuned to give the best synthesis performance. In our experiments, we use 0.9 for $ \omega_\text{mel}$, 1 for  $\omega_\text{post}$, 0.1 for $\omega_\text{end}$ and 0.9 for  $\omega_\text{bsc}$.

\section{Baseline: Modular Audiovisual Speech Synthesis} \label{tta_mod}

We use the standard Tacotron2 architecture and feed synthesized audio after reconstruction from predicted spectral features, e.g., using WaveRNN, as input for an audio-to-facial-animation module. 

We train a CNN-based audio-to-facial-animation model baseline. The network architecture is shown in Table \ref{table:AVFE}. The input acoustic features to this model is 40-dimensional filter bank features with a context window of 21 (10 past, current, and 10 future frames.) The acoustic features are extracted from the wav files. The acoustic speech synthesis is decoupled from the visual speech synthesis, so any text-to-speech system can be used. The ground truth BSCs described in Section \ref{sec:bsc} are used to train the network. The input to the penultimate fully connected layer is the concatenation of the former fully connected (FC) layer that encodes the audio features and the emotion embedding. 

In contrast to AVTacotron2, the audio-to-facial-animation model does not have access to the entire phoneme sequence. To compensate for this, the BSC loss $\mathcal{L}_\text{bsc}$ in the audio-to-facial-animation model is modified by weighting the loss for each entry in a batch differently. The weights of each entry are proportional to the importance of the corresponding spoken viseme (visual phoneme), so the network is penalized more for errors related to important visemes, such as /p/ and /w/ and less for errors related to less important visemes, such as /t/ and /g/. Table \ref{table:PHM} shows the viseme importance/weights, which are the classification accuracies of a video-based viseme classifier normalized over the maximum accuracy. The silence weight, which actually has the highest classification accuracy is set to zero. Using these weights improves the performance of the audio-to-facial-animation network. We use a hybrid loss of L1 and the cosine distance (see \cite{aldeneh2020self}), which was found to be more effective than the L1 and L2 for small, sparse, and bounded labels, such as the BSCs. To ensure fair comparison, we also experimented with the hybrid loss applied to Equation \ref{eq:total}, but found that L2 was easier to balance with the other losses with minimal impact on the overall performance measure.

\section{Experiments and Results}
\label{sec:results}

\subsection{Dataset}\label{sec:dataset}
\begin{table}[t!]
	\centering{
		\caption{The architecture of the audio-to-facial-animation network.}
		\label{table:AVFE}
			\begin{tabular}{cccccc}
				\hline
				Type 	&  \#Filters & Kernel & Stride/padd.  & 	Output  			& 		Act. \\
				\hline                                                                   
				Conv.  	&      128		& 3x3    &	1x1/same    & 		21x40x128		&    	RELU \\
				Conv.    	&	64		& 3x3    &	2x2/same    & 		11x20x64			&    	RELU \\
				Conv.    	&	64		& 3x3    &	1x1/same    & 		11x20x64		&    	RELU \\
				Conv.    	&	64		& 3x3    &	2x2/same    & 		6x10x64		&    	RELU \\
				Conv.    	&	64		& 3x3    &	1x1/valid    & 		4x8x64		&    	RELU \\
				Conv.    	&	64		& 3x3    &	1x1/valid    & 		2x6x64		&    	RELU \\
				FC    	&	512		& -        &	-   		  & 		1x1x512		&    	RELU \\
				FC+embed    	&	128		& -        &	-   		  & 		1x1x128		&    	RELU \\
				FC    	&	51		& -        &	-   		  & 		1x1x51		&    	None \\
			\hline                                                           
		\end{tabular}
	}
\end{table}
A  multimodal corpus was collected for training all of our models. The dataset consists of  10 hours of multimodal data captured from a professional actor: 7 hours of neutral speech ($\sim$7270 sentences), 1.5 hours ($\sim$1430) of acted happy speech and 1.5 hours ($\sim$1430) of acted sad speech. In this dataset, there are 74 minutes ($\sim$300) of common phrases spoken in each emotional state. The text corpus was balanced to maximize the phonetic coverage of US English phoneme-pairs. For each utterance a single channel 44.1kHz audio, 60 frame per second (fps) 1920x1080 RGB video, and  30 fps depth signals were captured. 

\subsection{Experimental  Setup and Results}\label{sec:exp_setup}

\subsubsection{Speech Emotion Recognition}\label{sec:emo_res}

For training the speech emotion recognition model, the corpus is randomly split into emotion-stratified partitions using a 80/10/10 rule. As described in Section \ref{sec:emotion_recog}, 40-dimensional MFB are used as input acoustic features. The network is trained to optimize the cross entropy loss using the Adam optimizer. We use a learning rate of 0.001 for a maximum of 20 epochs. We monitor the validation performance during training and apply early stopping when the validation loss converges. Since the data is class imbalanced, we penalize classification errors of each class inversely proportional to the class probability.

\begin{table}[t]
    \centering{
        \caption{The phoneme-viseme mapping and the viseme importance used to weight the loss of the audio-to-facial-animation-network}\label{table:PHM}
        \begin{tabular}{cccc}
                \midrule
                \textbf{Viseme cluster}              & \textbf{Viseme}                    &  \textbf{Phoneme}  &  \textbf{Weights}       \\
                \midrule
                 \toprule
                 Bilabial                             & /P/                       & /p/~/b/~/m/                 & 1.0\\
                 \midrule
                 Labio-Dental                         & /F/                       & /f/~/v/                     & 0.97\\
                 \midrule
                 \multirow{2}{*}{Palato alveolar}     & \multirow{2}{*}{/SH/}     & /sh/~/zh/~/ch/              & \multirow{2}{*}{0.75}\\
                                                     &                           & /jh/\\
                Dental                               & /TH/                      & /th/~/dh/                   & 0.66\\
                \midrule

                Alveolar fricative                   & /Z/                       & /z/~/s/                     & 0.66\\
                \midrule
                Lip rounded vowels                   &  \multirow{2}{*}{/V2/ }   & /uw/~/uh/~/ow/              & \multirow{2}{*}{0.6}\\
                level 2                              &                           & /w/ \\
                \midrule
                Lip rounded vowels                   &   \multirow{2}{*}{ /V1/ } & /aa/~/ah/~/ao/              & 0.59\\
                level 1                              &                           & /aw/~/er/~/oy/ \\
                \midrule
                Lip stretched vowels                 & \multirow{2}{*}{/V3/}     & /ae/~/eh/~/ey/              & \multirow{2}{*}{0.58}\\
                level 1                              &                           & /ay~/y/\\
                \midrule
                Alveolar semivowels                  & /L/                       & /l/~/el/~/r/                & 0.5\\
                \midrule
                Lip stretched vowels                 &  \multirow{2}{*}{/V4/}    & \multirow{2}{*}{/ih/~/iy/}  & \multirow{2}{*}{0.48}\\
                level 2                              &                           &    \\ 
                \midrule
                \multirow{2}{*}{Velar}               & \multirow{2}{*}{/G/}      & /g/~/ng/~/k/                & \multirow{2}{*}{0.46}\\
                                                     &                           & /hh/\\
                \midrule
                \multirow{2}{*}{Alveolar}            &\multirow{2}{*}{ /T/ }     & /t/~/d/~/n/                 & \multirow{2}{*}{0.36}\\
                                     &                           & /en/\\
                \midrule
                Silence                              & /SIL/                     & /sil/~/sp/                  & 0.0\\
                \midrule
        \end{tabular}
    }
\end{table} 

\begin{table}[t]
  \caption{Emotion classification performance from the Emotion Recognition System.}\label{tab:results_3}
  \centering
  \begin{tabular}{ccccc}
    \toprule
    \textbf{Emotion}  &\textbf{Precision (\%)}  &\textbf{Recall (\%)} &\textbf{F1-Score (\%)}  \\
    \midrule
      happy & 95.0 & 96.0 & 96.0  \\
      sad  & 99.0 & 100.0 & 99.0  \\
      neutral   & 99.0 & 99.0 & 99.0  \\
    \midrule
    \textbf{Accuracy (\%)}   & \multicolumn{3}{c}{\textbf{97.8}}  \\
    \bottomrule
  \end{tabular}
\label{tab:accu}
\end{table}

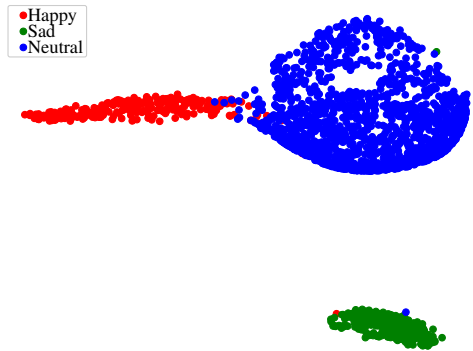
\begin{figure}[t!]
\centering
\def\svgwidth{0.6\linewidth}
\begingroup%
  \makeatletter%
  \providecommand\color[2][]{%
    \errmessage{(Inkscape) Color is used for the text in Inkscape, but the package 'color.sty' is not loaded}%
    \renewcommand\color[2][]{}%
  }%
  \providecommand\transparent[1]{%
    \errmessage{(Inkscape) Transparency is used (non-zero) for the text in Inkscape, but the package 'transparent.sty' is not loaded}%
    \renewcommand\transparent[1]{}%
  }%
  \providecommand\rotatebox[2]{#2}%
  \newcommand*\fsize{\dimexpr\f@size pt\relax}%
  \newcommand*\lineheight[1]{\fontsize{\fsize}{#1\fsize}\selectfont}%
  \ifx\svgwidth\undefined%
    \setlength{\unitlength}{575.9999856bp}%
    \ifx\svgscale\undefined%
      \relax%
    \else%
      \setlength{\unitlength}{\unitlength * \real{\svgscale}}%
    \fi%
  \else%
    \setlength{\unitlength}{\svgwidth}%
  \fi%
  \global\let\svgwidth\undefined%
  \global\let\svgscale\undefined%
  \makeatother%
  \begin{picture}(2.0,0.75)%
    \lineheight{1}%
    \setlength\tabcolsep{0pt}%
    \put(0.3,0){\includegraphics[width=\unitlength]{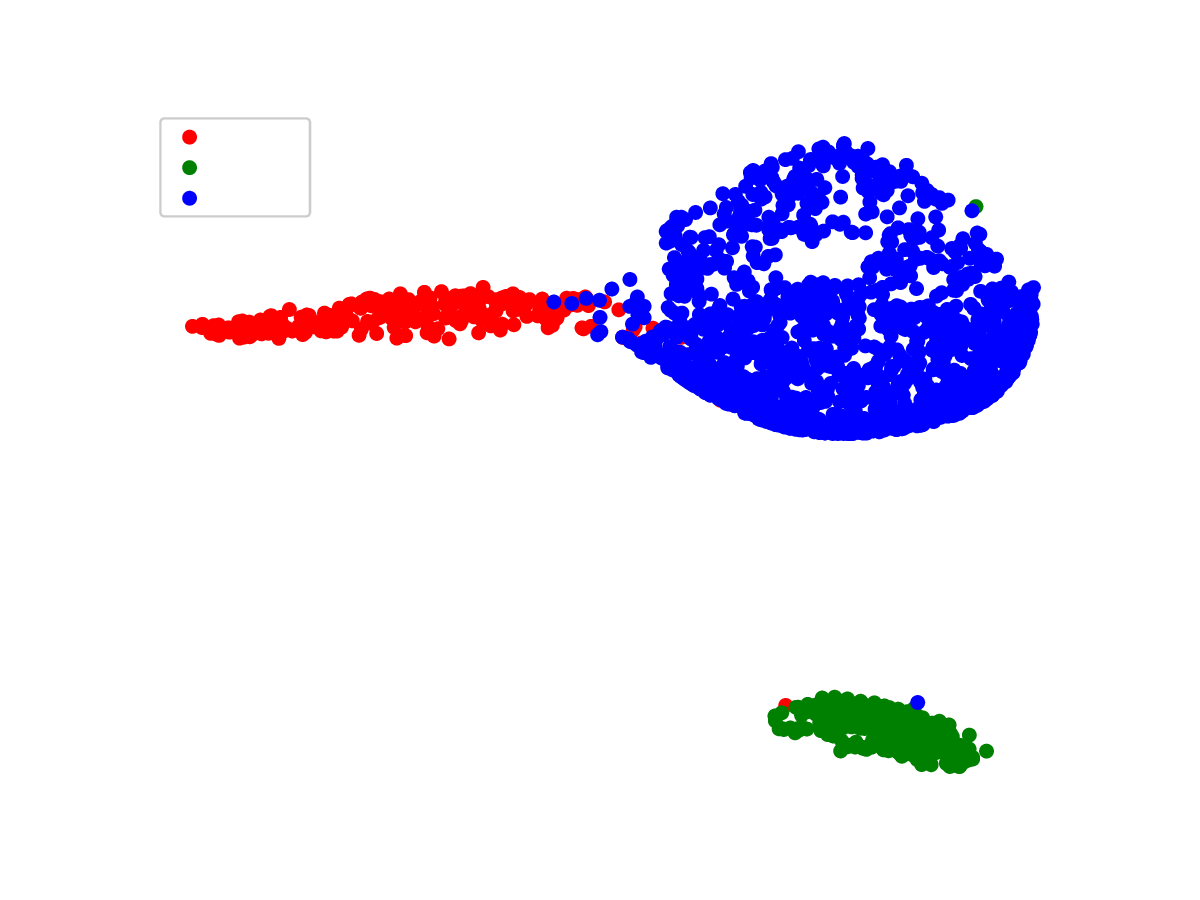}}%
    \put(0.465,0.627){\color[rgb]{0,0,0}\makebox(0,0)[lt]{\lineheight{1.25}\smash{\begin{tabular}[t]{l}\scriptsize Happy\end{tabular}}}}%
    \put(0.465,0.602){\color[rgb]{0,0,0}\makebox(0,0)[lt]{\lineheight{1.25}\smash{\begin{tabular}[t]{l}\scriptsize Sad\end{tabular}}}}%
    \put(0.465,0.575){\color[rgb]{0,0,0}\makebox(0,0)[lt]{\lineheight{1.25}\smash{\begin{tabular}[t]{l}\scriptsize Neutral\end{tabular}}}}%
  \end{picture}%
\endgroup%
\caption{TSNE projection of the validation set emotion embeddings extracted using the speech emotion recognition system.}
\label{fig:test3}
\end{figure}

Figure \ref{fig:test3} shows a TSNE projection \cite{van2008visualizing} of the emotion embeddings for the test set colored based on emotion labels. The emotion embeddings are well separated, which is reflected in the high classification accuracy shown in Table  \ref{tab:accu}. Although the performance of state-of-the-art speech emotion recognition systems in the wild does not reach such high accuracies \cite{tarantino2019self}, the high accuracy in Table \ref{tab:accu} was expected, as the emotions are acted. In contrast to simply using a one hot vector to represent the emotion state, the distance between emotion embeddings within the same cluster allows for fine-grained control of the emotion level that needs to be synthesized.

\subsubsection{Audiovisual Speech Synthesis}
Comparing machine learned models that produce synthesized audiovisual speech is challenging because objective measures, such as the loss functions used to train the network do not usually reflect the perceived naturalness. Human ratings of subjective quality are usually more reliable. In this study, we have conducted subjective tests to assess the quality of three different aspects of the synthesized talking face. The graders are gender balanced (15 male and 15 female) and they all are native US English speakers in the age range 21–50.

\textbf{Overall Experience of the Emotional Audiovisual Synthesis}\label{sec:vis_speech}
Graders are asked to evaluate videos from acoustic, visual, and overall naturalness perspectives. From the audio perspective, the graders are asked to evaluate the speech synthesis quality and emotion appropriateness given an emotion tag. To grade the visual aspects, the graders are asked to evaluate how much the visual synthesis (lip movements and facial expressions) match the synthesized speech. Finally, the graders are asked to evaluate the overall naturalness of the talking face. Mean opinion scores (MOSs) for the five questions described above ranging from (1) bad (no match) to (5) excellent (perfect match) are collected.

We used 30 videos, equally synthesized using happy and sad emotions. The MOS tests were conducted on each of the following sets of data --- videos generated by the modular, and end-to-end approaches and the original (ground truth) recordings. We also used the acoustic speech synthesized from the end-to-end approach to drive the audio-to-facial-animation module in the modular approach to ensure fair comparison. In this way the graders hear the same speech with different lip movements from the two systems. 

\textbf{Performance of feedback approaches in AVTacotron2}
We first conduct this test to evaluate the performance of the end-to-end models trained using full feedback and limited feedback approaches. The full feedback model achieved a higher overall MOS of 4.07 (compared to 3.9 from the limited feedback model). Both the models however achieve high MOS (>4) while evaluating just the facial expressions indicating that audio and lip movement features can be strong signals for synthesizing plausible facial expressions. In the tests below, we use the full feedback model for comparison with the modular approach baseline.

\begin{figure}[t!]
\centering
\def\svgwidth{0.6\linewidth}
\begingroup%
  \makeatletter%
  \providecommand\color[2][]{%
    \errmessage{(Inkscape) Color is used for the text in Inkscape, but the package 'color.sty' is not loaded}%
    \renewcommand\color[2][]{}%
  }%
  \providecommand\transparent[1]{%
    \errmessage{(Inkscape) Transparency is used (non-zero) for the text in Inkscape, but the package 'transparent.sty' is not loaded}%
    \renewcommand\transparent[1]{}%
  }%
  \providecommand\rotatebox[2]{#2}%
  \newcommand*\fsize{\dimexpr\f@size pt\relax}%
  \newcommand*\lineheight[1]{\fontsize{\fsize}{#1\fsize}\selectfont}%
  \ifx\svgwidth\undefined%
    \setlength{\unitlength}{438.87177133bp}%
    \ifx\svgscale\undefined%
      \relax%
    \else%
      \setlength{\unitlength}{\unitlength * \real{\svgscale}}%
    \fi%
  \else%
    \setlength{\unitlength}{\svgwidth}%
  \fi%
  \global\let\svgwidth\undefined%
  \global\let\svgscale\undefined%
  \makeatother%
  \begin{picture}(1,0.83)%
    \lineheight{1}%
    \setlength\tabcolsep{0pt}%
    \put(0,0){\includegraphics[width=\unitlength]{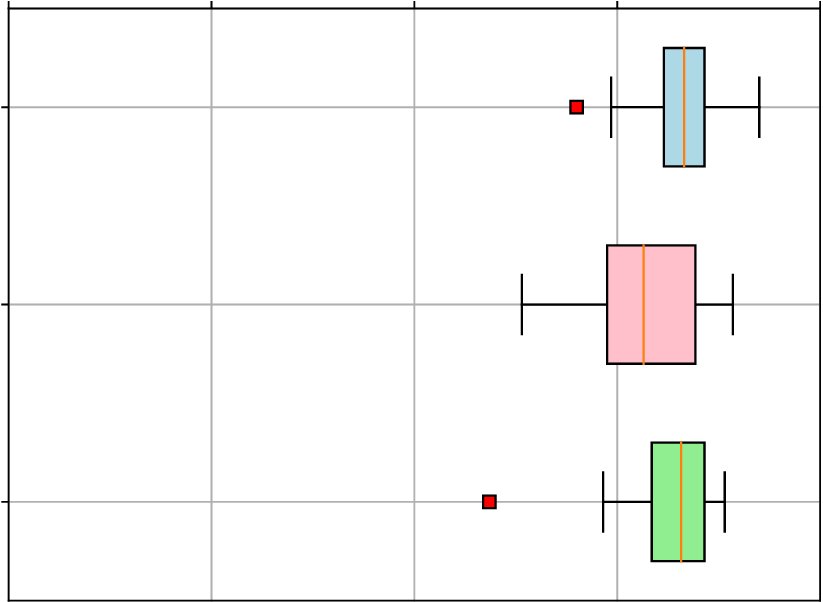}}%
    \put(-0.035,0.75){\color[rgb]{0,0,0}\makebox(0,0)[lt]{\lineheight{1.25}\smash{\begin{tabular}[t]{l}Bad\end{tabular}}}}%
    \put(0.21,0.75){\color[rgb]{0,0,0}\makebox(0,0)[lt]{\lineheight{1.25}\smash{\begin{tabular}[t]{l}Poor\end{tabular}}}}%
    \put(0.46438965,0.75){\color[rgb]{0,0,0}\makebox(0,0)[lt]{\lineheight{1.25}\smash{\begin{tabular}[t]{l}Fair\end{tabular}}}}%
    \put(0.69,0.75){\color[rgb]{0,0,0}\makebox(0,0)[lt]{\lineheight{1.25}\smash{\begin{tabular}[t]{l}Good\end{tabular}}}}%
    \put(0.9,0.75){\color[rgb]{0,0,0}\makebox(0,0)[lt]{\lineheight{1.25}\smash{\begin{tabular}[t]{l}Excellent\end{tabular}}}}%
    \put(-0.02,0.09){\color[rgb]{0,0,0}\rotatebox{90}{\makebox(0,0)[lt]{\lineheight{1.25}\smash{\begin{tabular}[t]{l}GT\end{tabular}}}}}%
    \put(0.84,0.15){\color[rgb]{0,0,0}\rotatebox{-90}{\makebox(0,0)[lt]{\lineheight{1.25}\smash{\begin{tabular}[t]{l}4.3\end{tabular}}}}}%
    \put(-0.02,0.29){\color[rgb]{0,0,0}\rotatebox{90}{\makebox(0,0)[lt]{\lineheight{1.25}\smash{\begin{tabular}[t]{l}Modular\end{tabular}}}}}%
    \put(0.79267685,0.39){\color[rgb]{0,0,0}\rotatebox{-90}{\makebox(0,0)[lt]{\lineheight{1.25}\smash{\begin{tabular}[t]{l}4.1\end{tabular}}}}}%
    \put(-0.02,0.56){\color[rgb]{0,0,0}\rotatebox{90}{\makebox(0,0)[lt]{\lineheight{1.25}\smash{\begin{tabular}[t]{l}E2E\end{tabular}}}}}%
     \put(0.84,0.627){\color[rgb]{0,0,0}\rotatebox{-90}{\makebox(0,0)[lt]{\lineheight{1.25}\smash{\begin{tabular}[t]{l}4.3\end{tabular}}}}}%
  \end{picture}%
\endgroup%
\caption{ Mean opinion scores of the lip movement quality in the videos generated by the end-to-end (E2E) and modular approaches, and the ground truth (GT) videos}
\label{fig:subj_MOS_overall_lip}
\end{figure}
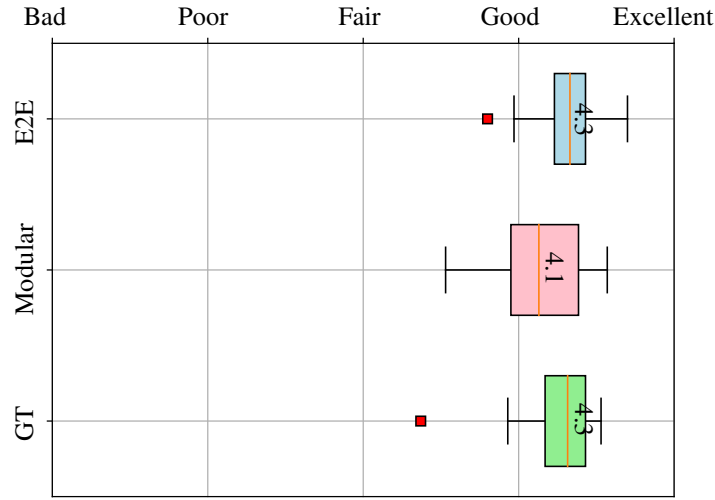

\begin{figure}[t!]
\centering
\def\svgwidth{0.6\linewidth}
\begingroup%
  \makeatletter%
  \providecommand\color[2][]{%
    \errmessage{(Inkscape) Color is used for the text in Inkscape, but the package 'color.sty' is not loaded}%
    \renewcommand\color[2][]{}%
  }%
  \providecommand\transparent[1]{%
    \errmessage{(Inkscape) Transparency is used (non-zero) for the text in Inkscape, but the package 'transparent.sty' is not loaded}%
    \renewcommand\transparent[1]{}%
  }%
  \providecommand\rotatebox[2]{#2}%
  \newcommand*\fsize{\dimexpr\f@size pt\relax}%
  \newcommand*\lineheight[1]{\fontsize{\fsize}{#1\fsize}\selectfont}%
  \ifx\svgwidth\undefined%
    \setlength{\unitlength}{438.87177133bp}%
    \ifx\svgscale\undefined%
      \relax%
    \else%
      \setlength{\unitlength}{\unitlength * \real{\svgscale}}%
    \fi%
  \else%
    \setlength{\unitlength}{\svgwidth}%
  \fi%
  \global\let\svgwidth\undefined%
  \global\let\svgscale\undefined%
  \makeatother%
  \begin{picture}(1,0.83)%
    \lineheight{1}%
    \setlength\tabcolsep{0pt}%
    \put(0,0){\includegraphics[width=\unitlength]{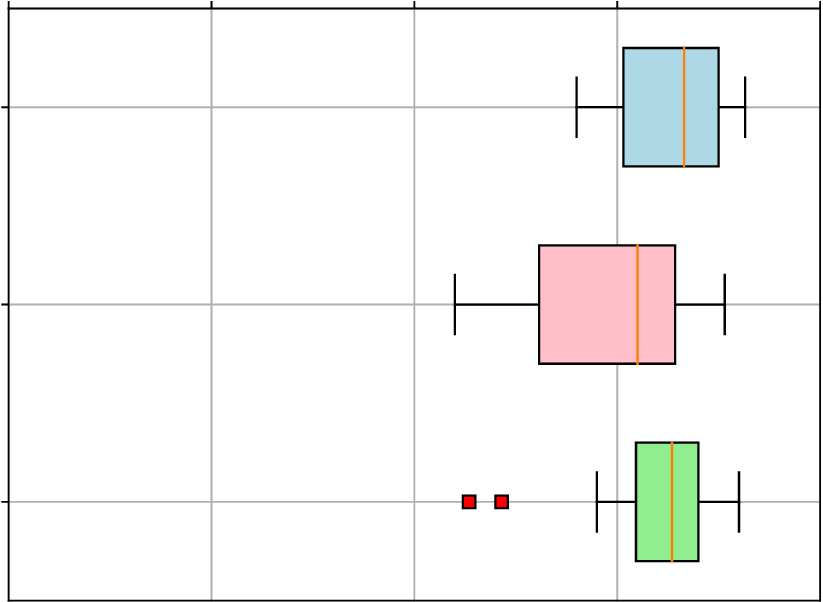}}%
    \put(-0.035,0.75){\color[rgb]{0,0,0}\makebox(0,0)[lt]{\lineheight{1.25}\smash{\begin{tabular}[t]{l}Bad\end{tabular}}}}%
    \put(0.21,0.75){\color[rgb]{0,0,0}\makebox(0,0)[lt]{\lineheight{1.25}\smash{\begin{tabular}[t]{l}Poor\end{tabular}}}}%
    \put(0.46438965,0.75){\color[rgb]{0,0,0}\makebox(0,0)[lt]{\lineheight{1.25}\smash{\begin{tabular}[t]{l}Fair\end{tabular}}}}%
    \put(0.69,0.75){\color[rgb]{0,0,0}\makebox(0,0)[lt]{\lineheight{1.25}\smash{\begin{tabular}[t]{l}Good\end{tabular}}}}%
    \put(0.9,0.75){\color[rgb]{0,0,0}\makebox(0,0)[lt]{\lineheight{1.25}\smash{\begin{tabular}[t]{l}Excellent\end{tabular}}}}%
    \put(-0.02,0.09){\color[rgb]{0,0,0}\rotatebox{90}{\makebox(0,0)[lt]{\lineheight{1.25}\smash{\begin{tabular}[t]{l}GT\end{tabular}}}}}%
    \put(0.84,0.15){\color[rgb]{0,0,0}\rotatebox{-90}{\makebox(0,0)[lt]{\lineheight{1.25}\smash{\begin{tabular}[t]{l}4.3\end{tabular}}}}}%
    \put(-0.02,0.29){\color[rgb]{0,0,0}\rotatebox{90}{\makebox(0,0)[lt]{\lineheight{1.25}\smash{\begin{tabular}[t]{l}Modular\end{tabular}}}}}%
    \put(0.79267685,0.39){\color[rgb]{0,0,0}\rotatebox{-90}{\makebox(0,0)[lt]{\lineheight{1.25}\smash{\begin{tabular}[t]{l}4.1\end{tabular}}}}}%
    \put(-0.02,0.56){\color[rgb]{0,0,0}\rotatebox{90}{\makebox(0,0)[lt]{\lineheight{1.25}\smash{\begin{tabular}[t]{l}E2E\end{tabular}}}}}%
     \put(0.84,0.627){\color[rgb]{0,0,0}\rotatebox{-90}{\makebox(0,0)[lt]{\lineheight{1.25}\smash{\begin{tabular}[t]{l}4.3\end{tabular}}}}}
  \end{picture}%
\endgroup%
\caption{Mean opinion scores of the facial expression quality in the videos generated by the end-to-end (E2E) and modular approaches and the ground truth (GT) videos.}
\label{fig:subj_MOS_fe}
\end{figure}
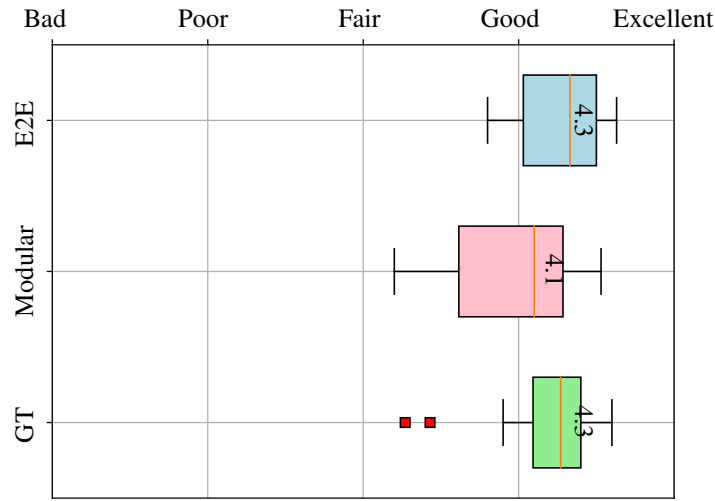

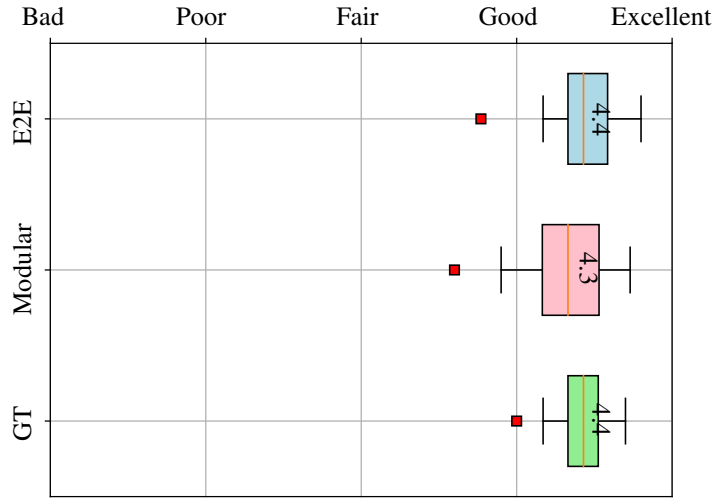
\begin{figure}[t!]
\centering
\def\svgwidth{0.6\linewidth}
\begingroup%
  \makeatletter%
  \providecommand\color[2][]{%
    \errmessage{(Inkscape) Color is used for the text in Inkscape, but the package 'color.sty' is not loaded}%
    \renewcommand\color[2][]{}%
  }%
  \providecommand\transparent[1]{%
    \errmessage{(Inkscape) Transparency is used (non-zero) for the text in Inkscape, but the package 'transparent.sty' is not loaded}%
    \renewcommand\transparent[1]{}%
  }%
  \providecommand\rotatebox[2]{#2}%
  \newcommand*\fsize{\dimexpr\f@size pt\relax}%
  \newcommand*\lineheight[1]{\fontsize{\fsize}{#1\fsize}\selectfont}%
  \ifx\svgwidth\undefined%
    \setlength{\unitlength}{438.87177133bp}%
    \ifx\svgscale\undefined%
      \relax%
    \else%
      \setlength{\unitlength}{\unitlength * \real{\svgscale}}%
    \fi%
  \else%
    \setlength{\unitlength}{\svgwidth}%
  \fi%
  \global\let\svgwidth\undefined%
  \global\let\svgscale\undefined%
  \makeatother%
  \begin{picture}(1,0.83)%
    \lineheight{1}%
    \setlength\tabcolsep{0pt}%
    \put(0,0){\includegraphics[width=\unitlength]{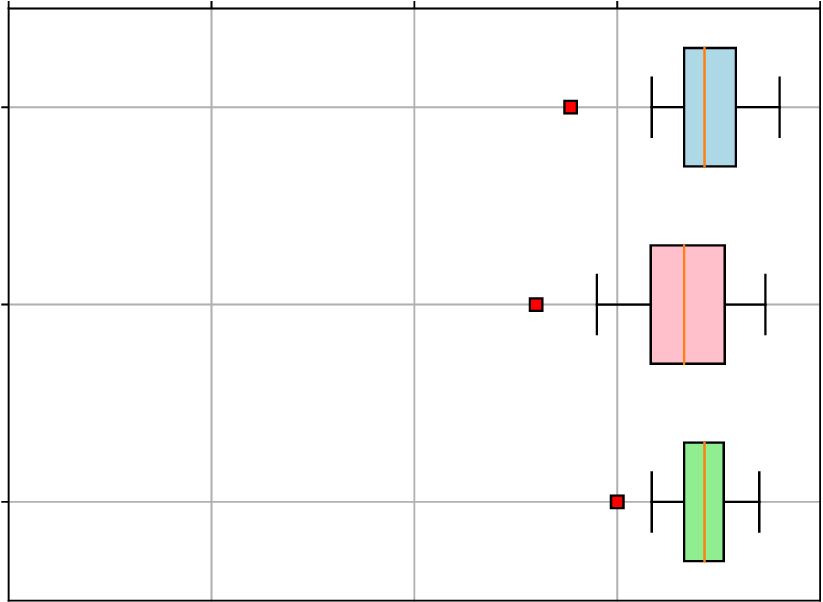}}%
    \put(-0.035,0.75){\color[rgb]{0,0,0}\makebox(0,0)[lt]{\lineheight{1.25}\smash{\begin{tabular}[t]{l}Bad\end{tabular}}}}%
    \put(0.21,0.75){\color[rgb]{0,0,0}\makebox(0,0)[lt]{\lineheight{1.25}\smash{\begin{tabular}[t]{l}Poor\end{tabular}}}}%
    \put(0.46438965,0.75){\color[rgb]{0,0,0}\makebox(0,0)[lt]{\lineheight{1.25}\smash{\begin{tabular}[t]{l}Fair\end{tabular}}}}%
    \put(0.69,0.75){\color[rgb]{0,0,0}\makebox(0,0)[lt]{\lineheight{1.25}\smash{\begin{tabular}[t]{l}Good\end{tabular}}}}%
    \put(0.9,0.75){\color[rgb]{0,0,0}\makebox(0,0)[lt]{\lineheight{1.25}\smash{\begin{tabular}[t]{l}Excellent\end{tabular}}}}%
    \put(-0.02,0.09){\color[rgb]{0,0,0}\rotatebox{90}{\makebox(0,0)[lt]{\lineheight{1.25}\smash{\begin{tabular}[t]{l}GT\end{tabular}}}}}%
    \put(0.87,0.15){\color[rgb]{0,0,0}\rotatebox{-90}{\makebox(0,0)[lt]{\lineheight{1.25}\smash{\begin{tabular}[t]{l}4.4\end{tabular}}}}}%
    \put(-0.02,0.29){\color[rgb]{0,0,0}\rotatebox{90}{\makebox(0,0)[lt]{\lineheight{1.25}\smash{\begin{tabular}[t]{l}Modular\end{tabular}}}}}%
    \put(0.85,0.39){\color[rgb]{0,0,0}\rotatebox{-90}{\makebox(0,0)[lt]{\lineheight{1.25}\smash{\begin{tabular}[t]{l}4.3\end{tabular}}}}}%
    \put(-0.02,0.56){\color[rgb]{0,0,0}\rotatebox{90}{\makebox(0,0)[lt]{\lineheight{1.25}\smash{\begin{tabular}[t]{l}E2E\end{tabular}}}}}%
     \put(0.87,0.627){\color[rgb]{0,0,0}\rotatebox{-90}{\makebox(0,0)[lt]{\lineheight{1.25}\smash{\begin{tabular}[t]{l}4.4\end{tabular}}}}}%
  \end{picture}%
\endgroup%
\caption{Mean opinion scores of acoustic speech quality synthesized by the end-to-end (E2E) approach to the ground truth (GT)}
\label{fig:subj_MOS_Syn}
\end{figure}

\begin{figure}[t!]
\centering
\def\svgwidth{0.6\linewidth}
\begingroup%
  \makeatletter%
  \providecommand\color[2][]{%
    \errmessage{(Inkscape) Color is used for the text in Inkscape, but the package 'color.sty' is not loaded}%
    \renewcommand\color[2][]{}%
  }%
  \providecommand\transparent[1]{%
    \errmessage{(Inkscape) Transparency is used (non-zero) for the text in Inkscape, but the package 'transparent.sty' is not loaded}%
    \renewcommand\transparent[1]{}%
  }%
  \providecommand\rotatebox[2]{#2}%
  \newcommand*\fsize{\dimexpr\f@size pt\relax}%
  \newcommand*\lineheight[1]{\fontsize{\fsize}{#1\fsize}\selectfont}%
  \ifx\svgwidth\undefined%
    \setlength{\unitlength}{438.87177133bp}%
    \ifx\svgscale\undefined%
      \relax%
    \else%
      \setlength{\unitlength}{\unitlength * \real{\svgscale}}%
    \fi%
  \else%
    \setlength{\unitlength}{\svgwidth}%
  \fi%
  \global\let\svgwidth\undefined%
  \global\let\svgscale\undefined%
  \makeatother%
  \begin{picture}(1,0.83)%
    \lineheight{1}%
    \setlength\tabcolsep{0pt}%
    \put(0,0){\includegraphics[width=\unitlength]{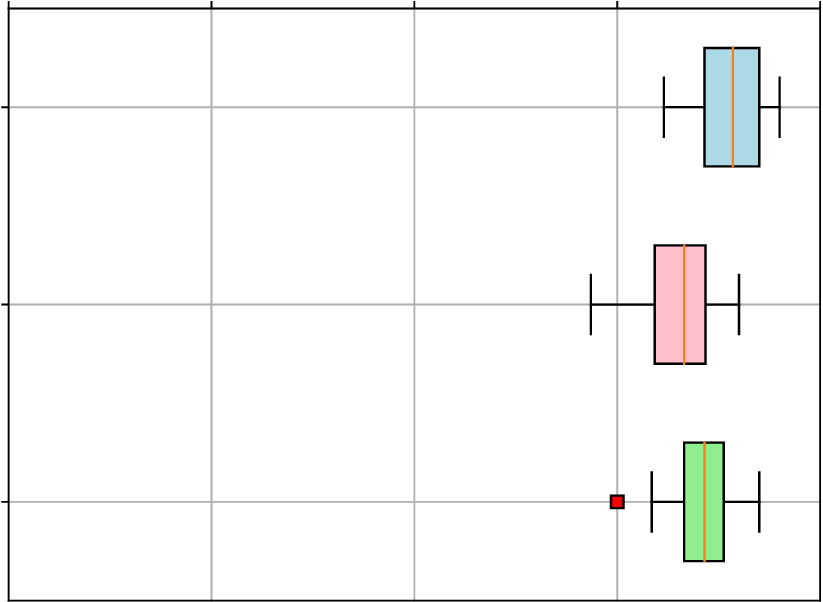}}%
    \put(-0.035,0.75){\color[rgb]{0,0,0}\makebox(0,0)[lt]{\lineheight{1.25}\smash{\begin{tabular}[t]{l}Bad\end{tabular}}}}%
    \put(0.21,0.75){\color[rgb]{0,0,0}\makebox(0,0)[lt]{\lineheight{1.25}\smash{\begin{tabular}[t]{l}Poor\end{tabular}}}}%
    \put(0.46438965,0.75){\color[rgb]{0,0,0}\makebox(0,0)[lt]{\lineheight{1.25}\smash{\begin{tabular}[t]{l}Fair\end{tabular}}}}%
    \put(0.69,0.75){\color[rgb]{0,0,0}\makebox(0,0)[lt]{\lineheight{1.25}\smash{\begin{tabular}[t]{l}Good\end{tabular}}}}%
    \put(0.9,0.75){\color[rgb]{0,0,0}\makebox(0,0)[lt]{\lineheight{1.25}\smash{\begin{tabular}[t]{l}Excellent\end{tabular}}}}%
    \put(-0.02,0.09){\color[rgb]{0,0,0}\rotatebox{90}{\makebox(0,0)[lt]{\lineheight{1.25}\smash{\begin{tabular}[t]{l}GT\end{tabular}}}}}%
    \put(0.87,0.15){\color[rgb]{0,0,0}\rotatebox{-90}{\makebox(0,0)[lt]{\lineheight{1.25}\smash{\begin{tabular}[t]{l}4.4\end{tabular}}}}}%
    \put(-0.02,0.29){\color[rgb]{0,0,0}\rotatebox{90}{\makebox(0,0)[lt]{\lineheight{1.25}\smash{\begin{tabular}[t]{l}Modular\end{tabular}}}}}%
    \put(0.84,0.39){\color[rgb]{0,0,0}\rotatebox{-90}{\makebox(0,0)[lt]{\lineheight{1.25}\smash{\begin{tabular}[t]{l}4.3\end{tabular}}}}}%
    \put(-0.02,0.56){\color[rgb]{0,0,0}\rotatebox{90}{\makebox(0,0)[lt]{\lineheight{1.25}\smash{\begin{tabular}[t]{l}E2E\end{tabular}}}}}%
     \put(0.9,0.627){\color[rgb]{0,0,0}\rotatebox{-90}{\makebox(0,0)[lt]{\lineheight{1.25}\smash{\begin{tabular}[t]{l}4.6\end{tabular}}}}}%
  \end{picture}%
\endgroup%
\caption{Mean opinion scores of the speech emotion quality synthesized by the end-to-end (E2E) to the ground truth (GT).}
\label{fig:subj_MOS_emo}
\end{figure}
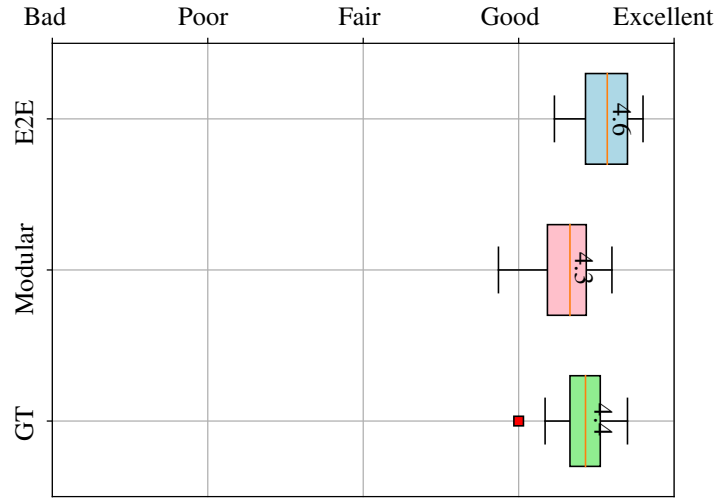

\begin{figure}[t!]
\centering
\def\svgwidth{0.6\linewidth}
\begingroup%
  \makeatletter%
  \providecommand\color[2][]{%
    \errmessage{(Inkscape) Color is used for the text in Inkscape, but the package 'color.sty' is not loaded}%
    \renewcommand\color[2][]{}%
  }%
  \providecommand\transparent[1]{%
    \errmessage{(Inkscape) Transparency is used (non-zero) for the text in Inkscape, but the package 'transparent.sty' is not loaded}%
    \renewcommand\transparent[1]{}%
  }%
  \providecommand\rotatebox[2]{#2}%
  \newcommand*\fsize{\dimexpr\f@size pt\relax}%
  \newcommand*\lineheight[1]{\fontsize{\fsize}{#1\fsize}\selectfont}%
  \ifx\svgwidth\undefined%
    \setlength{\unitlength}{438.87177133bp}%
    \ifx\svgscale\undefined%
      \relax%
    \else%
      \setlength{\unitlength}{\unitlength * \real{\svgscale}}%
    \fi%
  \else%
    \setlength{\unitlength}{\svgwidth}%
  \fi%
  \global\let\svgwidth\undefined%
  \global\let\svgscale\undefined%
  \makeatother%
  \begin{picture}(1,0.83)%
    \lineheight{1}%
    \setlength\tabcolsep{0pt}%
    \put(0,0){\includegraphics[width=\unitlength]{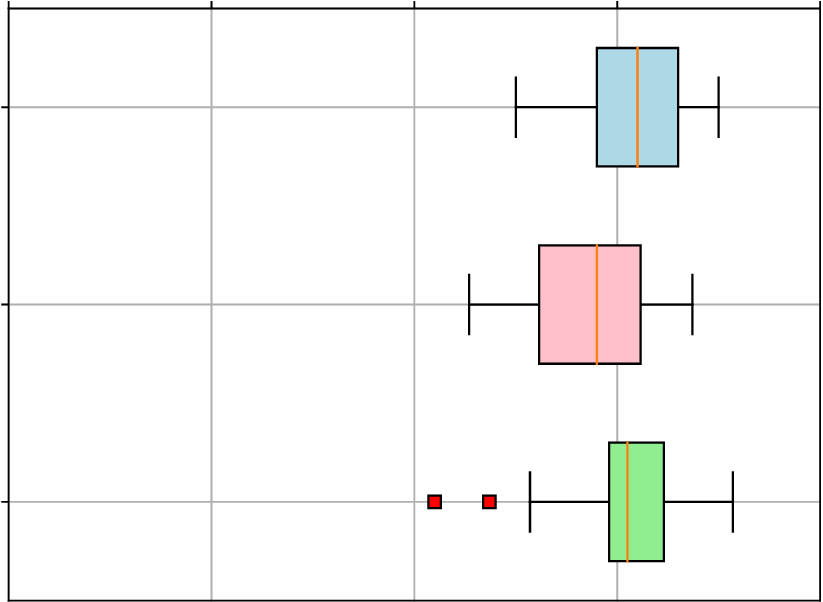}}%
    \put(-0.035,0.75){\color[rgb]{0,0,0}\makebox(0,0)[lt]{\lineheight{1.25}\smash{\begin{tabular}[t]{l}Bad\end{tabular}}}}%
    \put(0.21,0.75){\color[rgb]{0,0,0}\makebox(0,0)[lt]{\lineheight{1.25}\smash{\begin{tabular}[t]{l}Poor\end{tabular}}}}%
    \put(0.46438965,0.75){\color[rgb]{0,0,0}\makebox(0,0)[lt]{\lineheight{1.25}\smash{\begin{tabular}[t]{l}Fair\end{tabular}}}}%
    \put(0.69,0.75){\color[rgb]{0,0,0}\makebox(0,0)[lt]{\lineheight{1.25}\smash{\begin{tabular}[t]{l}Good\end{tabular}}}}%
    \put(0.9,0.75){\color[rgb]{0,0,0}\makebox(0,0)[lt]{\lineheight{1.25}\smash{\begin{tabular}[t]{l}Excellent\end{tabular}}}}%
    \put(-0.02,0.09){\color[rgb]{0,0,0}\rotatebox{90}{\makebox(0,0)[lt]{\lineheight{1.25}\smash{\begin{tabular}[t]{l}GT\end{tabular}}}}}%
    \put(0.78,0.15){\color[rgb]{0,0,0}\rotatebox{-90}{\makebox(0,0)[lt]{\lineheight{1.25}\smash{\begin{tabular}[t]{l}4.1\end{tabular}}}}}%
    \put(-0.02,0.29){\color[rgb]{0,0,0}\rotatebox{90}{\makebox(0,0)[lt]{\lineheight{1.25}\smash{\begin{tabular}[t]{l}Modular\end{tabular}}}}}%
    \put(0.74,0.39){\color[rgb]{0,0,0}\rotatebox{-90}{\makebox(0,0)[lt]{\lineheight{1.25}\smash{\begin{tabular}[t]{l}3.9\end{tabular}}}}}%
    \put(-0.02,0.56){\color[rgb]{0,0,0}\rotatebox{90}{\makebox(0,0)[lt]{\lineheight{1.25}\smash{\begin{tabular}[t]{l}E2E\end{tabular}}}}}%
     \put(0.788,0.627){\color[rgb]{0,0,0}\rotatebox{-90}{\makebox(0,0)[lt]{\lineheight{1.25}\smash{\begin{tabular}[t]{l}4.1\end{tabular}}}}}%
  \end{picture}%
\endgroup%
\caption{Mean opinion scores of the overall quality of the emotional talking faces in the videos generated by the end-to-end (E2E) and modular approaches, and the ground truth (GT) videos.}
\label{fig:subj_MOS_oa}
\end{figure}
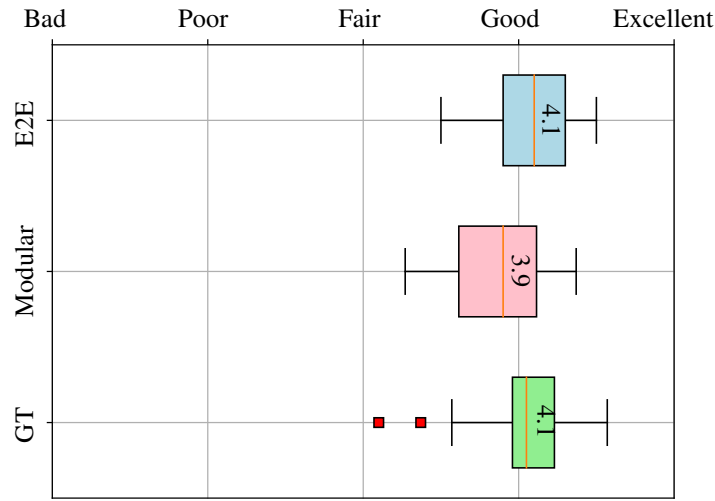

\begin{figure}[t!]
\centering
\def\svgwidth{0.6\linewidth}
\begingroup%
  \makeatletter%
  \providecommand\color[2][]{%
    \errmessage{(Inkscape) Color is used for the text in Inkscape, but the package 'color.sty' is not loaded}%
    \renewcommand\color[2][]{}%
  }%
  \providecommand\transparent[1]{%
    \errmessage{(Inkscape) Transparency is used (non-zero) for the text in Inkscape, but the package 'transparent.sty' is not loaded}%
    \renewcommand\transparent[1]{}%
  }%
  \providecommand\rotatebox[2]{#2}%
  \newcommand*\fsize{\dimexpr\f@size pt\relax}%
  \newcommand*\lineheight[1]{\fontsize{\fsize}{#1\fsize}\selectfont}%
  \ifx\svgwidth\undefined%
    \setlength{\unitlength}{438.87176228bp}%
    \ifx\svgscale\undefined%
      \relax%
    \else%
      \setlength{\unitlength}{\unitlength * \real{\svgscale}}%
    \fi%
  \else%
    \setlength{\unitlength}{\svgwidth}%
  \fi%
  \global\let\svgwidth\undefined%
  \global\let\svgscale\undefined%
  \makeatother%
  \begin{picture}(1,0.83)%
    \lineheight{1}%
    \setlength\tabcolsep{0pt}%
    \put(0,0){\includegraphics[width=\unitlength]{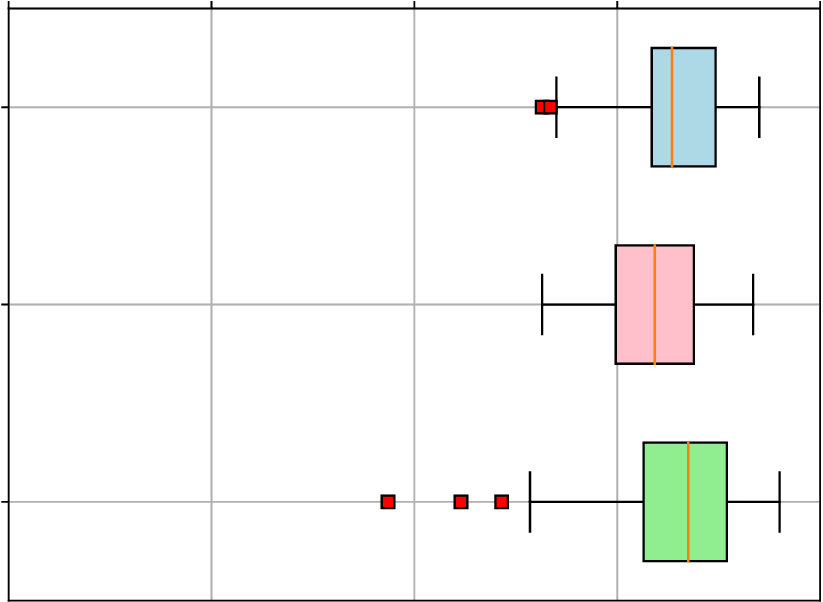}}%
    \put(-0.035,0.75){\color[rgb]{0,0,0}\makebox(0,0)[lt]{\lineheight{1.25}\smash{\begin{tabular}[t]{l}Bad\end{tabular}}}}%
    \put(0.21,0.75){\color[rgb]{0,0,0}\makebox(0,0)[lt]{\lineheight{1.25}\smash{\begin{tabular}[t]{l}Poor\end{tabular}}}}%
    \put(0.46438965,0.75){\color[rgb]{0,0,0}\makebox(0,0)[lt]{\lineheight{1.25}\smash{\begin{tabular}[t]{l}Fair\end{tabular}}}}%
    \put(0.69,0.75){\color[rgb]{0,0,0}\makebox(0,0)[lt]{\lineheight{1.25}\smash{\begin{tabular}[t]{l}Good\end{tabular}}}}%
    \put(0.9,0.75){\color[rgb]{0,0,0}\makebox(0,0)[lt]{\lineheight{1.25}\smash{\begin{tabular}[t]{l}Excellent\end{tabular}}}}%
    \put(-0.02,0.09){\color[rgb]{0,0,0}\rotatebox{90}{\makebox(0,0)[lt]{\lineheight{1.25}\smash{\begin{tabular}[t]{l}GT\end{tabular}}}}}%
    \put(0.85,0.15){\color[rgb]{0,0,0}\rotatebox{-90}{\makebox(0,0)[lt]{\lineheight{1.25}\smash{\begin{tabular}[t]{l}4.4\end{tabular}}}}}%
    \put(-0.02,0.29){\color[rgb]{0,0,0}\rotatebox{90}{\makebox(0,0)[lt]{\lineheight{1.25}\smash{\begin{tabular}[t]{l}Modular\end{tabular}}}}}%
    \put(0.805,0.39){\color[rgb]{0,0,0}\rotatebox{-90}{\makebox(0,0)[lt]{\lineheight{1.25}\smash{\begin{tabular}[t]{l}4.2\end{tabular}}}}}%
    \put(-0.02,0.56){\color[rgb]{0,0,0}\rotatebox{90}{\makebox(0,0)[lt]{\lineheight{1.25}\smash{\begin{tabular}[t]{l}E2E\end{tabular}}}}}%
     \put(0.83,0.627){\color[rgb]{0,0,0}\rotatebox{-90}{\makebox(0,0)[lt]{\lineheight{1.25}\smash{\begin{tabular}[t]{l}4.3\end{tabular}}}}}%
  \end{picture}%
\endgroup%
\caption{Mean opinion scores of the visual speech synthesis of the end-to-end (E2E) and the modular approaches in  isolated setup.}
\label{fig:subj_MOS_lip}
\end{figure}
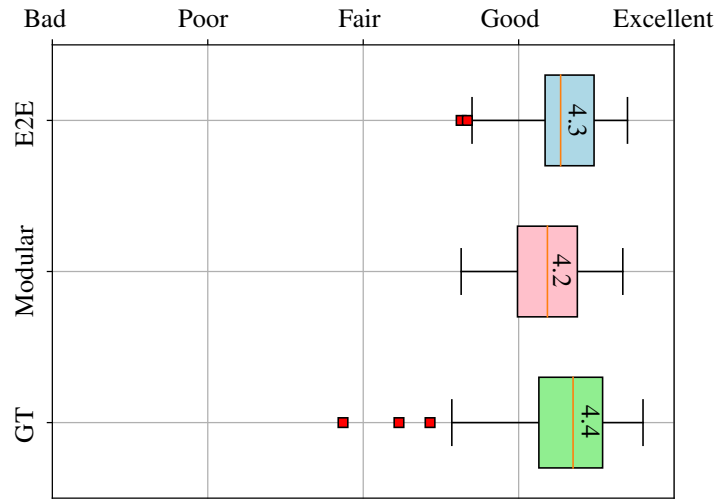

For all tests, the graders gave high scores (between good and excellent) as shown in Figures \ref{fig:subj_MOS_overall_lip}-\ref{fig:subj_MOS_oa}. All results are significant according to the Mann-Whitney significance test. Figures \ref{fig:subj_MOS_overall_lip}-\ref{fig:subj_MOS_oa} show that AVTacotron2 gives better and more consistent performance, i.e, with small variance, compared to the modular approach in all aspects. Furthermore, the performance of AVTacotron2 is close to the ground truth results. Figures \ref{fig:subj_MOS_overall_lip}-\ref{fig:subj_MOS_fe}  show results of grading the visual aspects of the synthesized talking face, i.e, lip movements and facial expressions. The ground truth results depicted in Figures \ref{fig:subj_MOS_overall_lip}-\ref{fig:subj_MOS_fe} are also an indication of the performance of the offline blendshape estimation algorithms discussed in Section \ref{sec:bsc}. The ground truth BSCs are estimated from RGB+depth images and the quality of these inputs relies on many factors, such as the head pose and lighting conditions. The outliers in the ground truth results in Figures \ref{fig:subj_MOS_overall_lip}, \ref{fig:subj_MOS_fe} and \ref{fig:subj_MOS_lip} are explained by cases where the BSC estimation may have failed when such conditions are not optimal. Neural networks on the other hand, when trained on these data, learn to filter out such outliers and infer the conditional mean. Therefore, AVTacotron2 shows a more consistent performance in Figures \ref{fig:subj_MOS_overall_lip}-\ref{fig:subj_MOS_fe}. In Figure \ref{fig:subj_MOS_fe}, the larger variance of the modular synthesis approach is explained by the audio-to-animation model falling short in synthesizing reasonable emotional facial expressions from speech.

An interesting outcome from this study was that although the graders were instructed to completely ignore visual aspects when they grade the acoustic speech synthesis, they seemed to be subconsciously affected by the visual aspects. This effect can be seen in the evaluation of the acoustic speech synthesis in Figures \ref{fig:subj_MOS_Syn} and \ref{fig:subj_MOS_emo}. Although, the speech synthesized from AVTacotron2 was the same speech used as input to the audio-to-animation model, the graders gave better speech synthesis and emotion scores to talking faces generated using AVTacotron2.

Another interesting result is shown in Figure \ref{fig:subj_MOS_emo}, where graders gave higher scores for the emotional speech synthesized from AVTacotron2 over the original recordings (GT). The reason for this is that in the original recording, the actor exaggerated happy and sad emotions in certain cases. We noticed that the emotions synthesized from AVTacotron2 are less exaggerated than the original recordings. This effect is discussed further in \ref{sec:vis_speech}. The graders preferred less exaggerated happy and sad emotions in comparison to the overacted ones in the original dataset. 

To evaluate the performance of AVTacotron2 and the modular approaches in terms of acoustic and visual speech synthesis in a more independent manner, we introduce two additional subjective tests described next to evaluate visual and acoustic speech synthesis performance isolated from all other aspects. 

\textbf{Visual Speech Synthesis}\label{sec:res_vis_speech}

We conducted an AB test to compare the performance of AVTacotron2 and the modular approach in terms of the synthesized lip movements (visual speech), i.e. how naturally the avatar's lip movements followed speech. To make sure the graders focus only on the quality of the synthesized visual speech, we used a neutral emotion embedding to synthesize neutral speech. We also eliminated (zeroed out) the controls of the upper half of the face, such as eyebrow movements and blinking. Human graders were given 50 pairs of videos that were not used during training. The graders were asked \textit{which video matches the speech more naturally”?} Graders were advised to focus only on the lip movements and to provide comments as to why they prefer their chosen video. To prevent display ordering effects, the order of the videos in a given pair is randomized. In total, 30 graders evaluated 50 videos. 

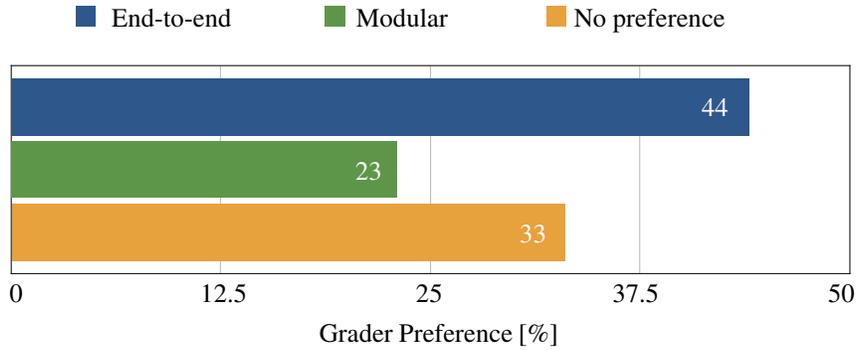
\begin{figure}[t!]
\centering
\def\svgwidth{0.8\linewidth}
\begingroup%
  \makeatletter%
  \providecommand\color[2][]{%
    \errmessage{(Inkscape) Color is used for the text in Inkscape, but the package 'color.sty' is not loaded}%
    \renewcommand\color[2][]{}%
  }%
  \providecommand\transparent[1]{%
    \errmessage{(Inkscape) Transparency is used (non-zero) for the text in Inkscape, but the package 'transparent.sty' is not loaded}%
    \renewcommand\transparent[1]{}%
  }%
  \providecommand\rotatebox[2]{#2}%
  \newcommand*\fsize{\dimexpr\f@size pt\relax}%
  \newcommand*\lineheight[1]{\fontsize{\fsize}{#1\fsize}\selectfont}%
  \ifx\svgwidth\undefined%
    \setlength{\unitlength}{850.61597336bp}%
    \ifx\svgscale\undefined%
      \relax%
    \else%
      \setlength{\unitlength}{\unitlength * \real{\svgscale}}%
    \fi%
  \else%
    \setlength{\unitlength}{\svgwidth}%
  \fi%
  \global\let\svgwidth\undefined%
  \global\let\svgscale\undefined%
  \makeatother%
  \begin{picture}(1,0.35362984)%
    \lineheight{1}%
    \setlength\tabcolsep{0pt}%
    \put(0,0){\includegraphics[width=\unitlength]{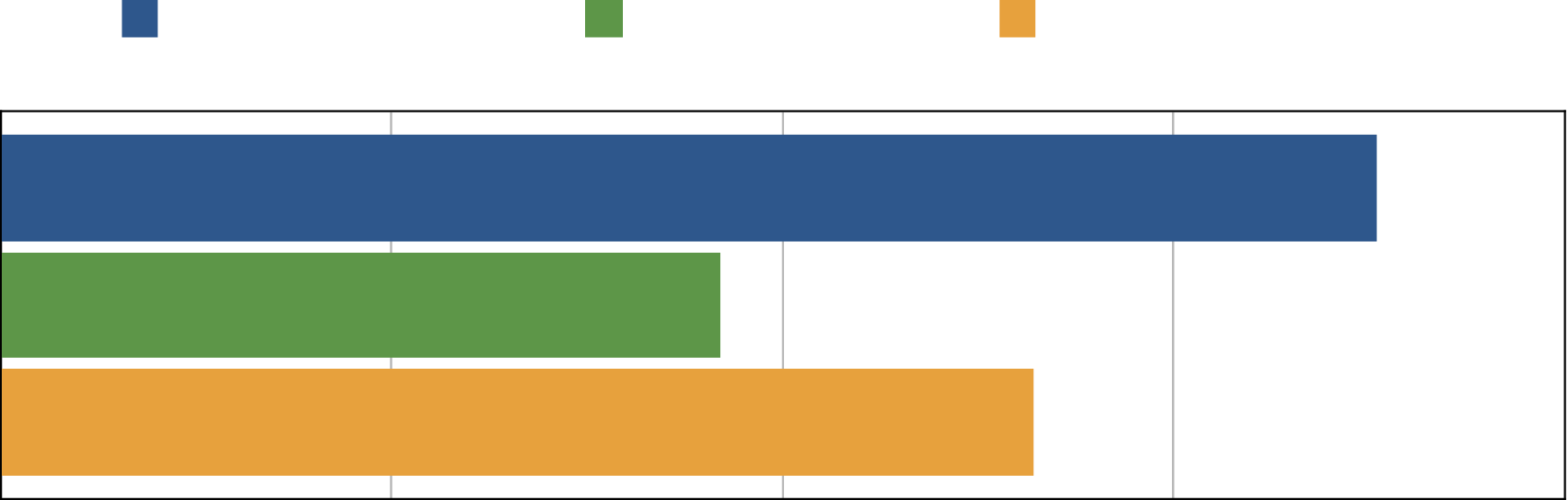}}%
    \put(0.36688236,-0.081){\color[rgb]{0,0,0}\makebox(0,0)[lt]{\lineheight{1.25}\smash{\begin{tabular}[t]{l}Grader Preference [\%] \end{tabular}}}}%
    \put(-0.00108627,-0.034){\color[rgb]{0,0,0}\makebox(0,0)[lt]{\lineheight{1.25}\smash{\begin{tabular}[t]{l}0\end{tabular}}}}%
    \put(0.22580812,-0.034){\color[rgb]{0,0,0}\makebox(0,0)[lt]{\lineheight{1.25}\smash{\begin{tabular}[t]{l}12.5\end{tabular}}}}%
    \put(0.48209298,-0.034){\color[rgb]{0,0,0}\makebox(0,0)[lt]{\lineheight{1.25}\smash{\begin{tabular}[t]{l}25\end{tabular}}}}%
    \put(0.71604109,-0.034){\color[rgb]{0,0,0}\makebox(0,0)[lt]{\lineheight{1.25}\smash{\begin{tabular}[t]{l}37.5\end{tabular}}}}%
    \put(0.97232594,-0.034){\color[rgb]{0,0,0}\makebox(0,0)[lt]{\lineheight{1.25}\smash{\begin{tabular}[t]{l}50\end{tabular}}}}%
    \put(0.60553294,0.038){\color[rgb]{1,1,1}\makebox(0,0)[lt]{\lineheight{1.25}\smash{\begin{tabular}[t]{l}33\end{tabular}}}}%
    \put(0.41038024,0.112){\color[rgb]{1,1,1}\makebox(0,0)[lt]{\lineheight{1.25}\smash{\begin{tabular}[t]{l}23\end{tabular}}}}%
    \put(0.82184676,0.188){\color[rgb]{1,1,1}\makebox(0,0)[lt]{\lineheight{1.25}\smash{\begin{tabular}[t]{l}44\end{tabular}}}}%
    \put(0.12,0.295){\color[rgb]{0,0,0}\makebox(0,0)[lt]{\lineheight{1.25}\smash{\begin{tabular}[t]{l}End-to-end\end{tabular}}}}%
    \put(0.411,0.295){\color[rgb]{0,0,0}\makebox(0,0)[lt]{\lineheight{1.25}\smash{\begin{tabular}[t]{l}Modular\end{tabular}}}}%
    \put(0.670,0.295){\color[rgb]{0,0,0}\makebox(0,0)[lt]{\lineheight{1.25}\smash{\begin{tabular}[t]{l}No preference\end{tabular}}}}%
  \end{picture}%
\endgroup%
\vspace{10mm}
\caption{Results of the ABX test that comparing the quality of the avatar's lip movements that are driven by the end-to-end and the modular approaches in an isolated setup.}
\label{fig:subj_ABX}
\end{figure}

Figure \ref{fig:subj_ABX} shows that the graders preferred the performance of AVTacotron2 more than the modular one. This result is significant according to Mann-Whitney significance test. The binary result of the AB test does not assess the \emph{absolute} quality of the visual speech synthesized by any of the two approaches. To get an absolute quality score for each of the two systems, we conducted two additional five point scale MOS tests for each approach independently using a setup similar to the AB test. The MOS test results shown in Figure \ref{fig:subj_MOS_lip} shows that while AVTacotron2 is better than the modular approach, both systems generate high quality lip movements. 

\textbf{7.2.2.3 Acoustic Speech Synthesis}\label{sec:vis_speech}
It is important to assess whether the quality of the synthesized acoustic speech is affected by adding the facial control estimation task to the Tacotron2 model. To assess this, we conducted audio-only MOS tests for neutral speech, where graders were asked to primarily focus on assessing the quality of the synthesis. The graders were given 50 pairs of audio signals. Each pair consisted of a reference recording from the actor and a synthesized audio signal of the same phrase. The graders were asked to evaluate the synthesis quality given the reference signals on a five point scale, where one corresponds to poor quality and five to excellent quality. The MOS test was conducted first with the audiovisual Tacotron2 and next with the audio-only Tacotron2. 

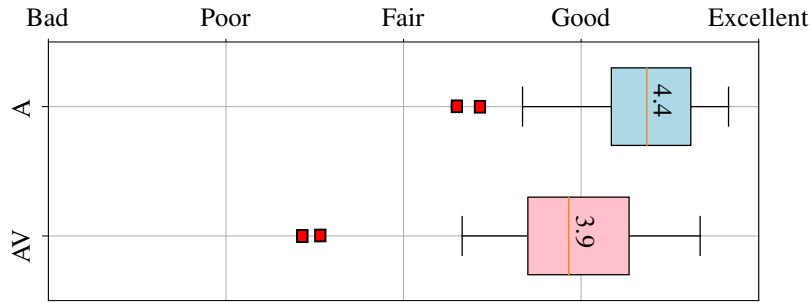
\begin{figure}[t!]
\centering
\def\svgwidth{0.8\linewidth}
\begingroup%
  \makeatletter%
  \providecommand\color[2][]{%
    \errmessage{(Inkscape) Color is used for the text in Inkscape, but the package 'color.sty' is not loaded}%
    \renewcommand\color[2][]{}%
  }%
  \providecommand\transparent[1]{%
    \errmessage{(Inkscape) Transparency is used (non-zero) for the text in Inkscape, but the package 'transparent.sty' is not loaded}%
    \renewcommand\transparent[1]{}%
  }%
  \providecommand\rotatebox[2]{#2}%
  \newcommand*\fsize{\dimexpr\f@size pt\relax}%
  \newcommand*\lineheight[1]{\fontsize{\fsize}{#1\fsize}\selectfont}%
  \ifx\svgwidth\undefined%
    \setlength{\unitlength}{460.39998849bp}%
    \ifx\svgscale\undefined%
      \relax%
    \else%
      \setlength{\unitlength}{\unitlength * \real{\svgscale}}%
    \fi%
  \else%
    \setlength{\unitlength}{\svgwidth}%
  \fi%
  \global\let\svgwidth\undefined%
  \global\let\svgscale\undefined%
  \makeatother%
  \begin{picture}(1,0.45)%
    \lineheight{1}%
    \setlength\tabcolsep{0pt}%
    \put(0,0){\includegraphics[width=\unitlength]{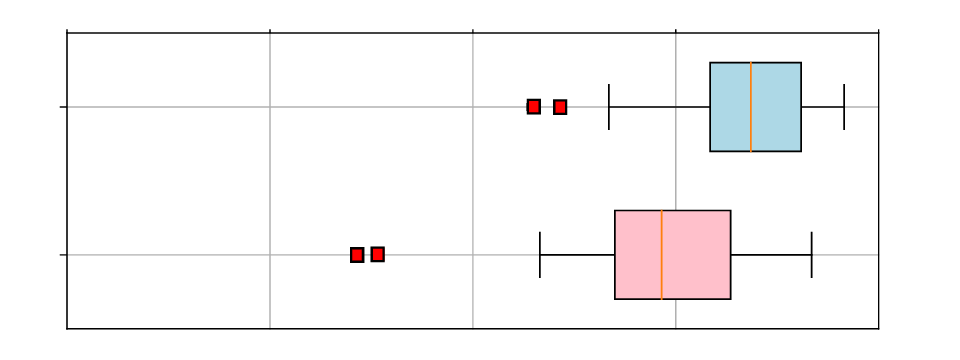}}%
    \put(0.04370526,0.36){\color[rgb]{0,0,0}\makebox(0,0)[lt]{\lineheight{1.25}\smash{\begin{tabular}[t]{l}Bad\end{tabular}}}}%
    \put(0.25067984,0.36){\color[rgb]{0,0,0}\makebox(0,0)[lt]{\lineheight{1.25}\smash{\begin{tabular}[t]{l}Poor\end{tabular}}}}%
    \put(0.46673328,0.36){\color[rgb]{0,0,0}\makebox(0,0)[lt]{\lineheight{1.25}\smash{\begin{tabular}[t]{l}Fair\end{tabular}}}}%
    \put(0.66831234,0.36){\color[rgb]{0,0,0}\makebox(0,0)[lt]{\lineheight{1.25}\smash{\begin{tabular}[t]{l}Good\end{tabular}}}}%
    \put(0.85451781,0.36){\color[rgb]{0,0,0}\makebox(0,0)[lt]{\lineheight{1.25}\smash{\begin{tabular}[t]{l}Excellent\end{tabular}}}}%
    \put(0.0477778,0.077){\color[rgb]{0,0,0}\rotatebox{90}{\makebox(0,0)[lt]{\lineheight{1.25}\smash{\begin{tabular}[t]{l}AV\end{tabular}}}}}%
    \put(0.0477778,0.25){\color[rgb]{0,0,0}\rotatebox{90}{\makebox(0,0)[lt]{\lineheight{1.25}\smash{\begin{tabular}[t]{l}A\end{tabular}}}}}%
    \put(0.69823414,0.135){\color[rgb]{0,0,0}\rotatebox{-90}{\makebox(0,0)[lt]{\lineheight{1.25}\smash{\begin{tabular}[t]{l}3.9\end{tabular}}}}}%
    \put(0.79,0.29){\color[rgb]{0,0,0}\rotatebox{-90}{\makebox(0,0)[lt]{\lineheight{1.25}\smash{\begin{tabular}[t]{l}4.4\end{tabular}}}}}%
  \end{picture}%
\endgroup%
\caption{Mean opinion scores of the acoustic speech synthesis of the audiovisual (AV) and audio-only (A) synthesis systems that use Tacotron2 architecture.}
\label{fig:subj_MOS_tts}
\end{figure}

The results of the two MOS tests are shown in Figure \ref{fig:subj_MOS_tts}. These results are significant according to Mann-Whitney significance test. Although the graders were explicitly instructed to focus only on the synthesis quality and not on the style of speech in the synthesized signal, the results and subsequent comments from the graders indicated that they were subconsciously affected by the speaking style. The low MOSs of the AVTacotron2 in Figure \ref{fig:subj_MOS_tts} is heavily weighted by the fact that the prosody of the synthesized speech does not always match the prosody of the reference signal. As shown in Figure \ref{fig:subj_MOS_tts}, this was not the case in the traditional audio-only Tacotron2,  where the prosody of the synthesis matches that of the reference signals and results in a higher MOS. This was also the reason why the graders preferred the emotional speech synthesis of AVTacotron2 than the overacted original recordings. 

\section{Conclusion} \label{sec:conclsion}
In this paper, we presented AVTacotron2, an end-to-end audiovisual speech synthesizer that generates both acoustic and visual speech from text. We compared against a baseline modular system that uses the conventional audio-only Tacotron2 to synthesize acoustic speech from text. The synthesized speech is then used as input to a speech-to-animation module that generates the corresponding facial controls. Emotion embeddings from a speech emotion classifier are used to condition both systems to synthesize emotional audiovisual speech. Both approaches are evaluated using subjective evaluation. The results show that AVTacotron2 outperforms the modular approach in all aspects. The additional task of facial control estimation in the end-to-end approach can sometimes lead to a mismatch between the prosody of the synthesized acoustic speech and the original recordings. AVTacotron2 produces synthesized talking faces that are rated almost as natural as the ground truth and can be applied to a variety of fictional and human-like face models without post-processing. 
To enhance the experience of the talking faces, our on-going work involves the task of estimating head pose. Finally, we are investigating video-based and audiovisual-based emotion embeddings compared to the audio-only one used here.

\section{Acknowledgments}
The authors are grateful to Barry-John Theobald, Oggi Rudovic, and Javier Latorre Martinez for their valuable comments.

\bibliographystyle{plain}
\bibliography{bibliography}

\end{document}